\documentclass{article}

    \PassOptionsToPackage{numbers, compress}{natbib}

\usepackage[preprint]{neurips_2020}

\usepackage{graphicx}
\usepackage{comment}
\usepackage{amsmath,amssymb} 
\usepackage{color}
\usepackage[utf8]{inputenc} 
\usepackage[T1]{fontenc}    
\usepackage{hyperref}       
\usepackage{url}            
\usepackage{booktabs}       
\usepackage{amsfonts}       
\usepackage{nicefrac}       
\usepackage{enumerate}
\usepackage{microtype}      

\title{\textit{fAshIon} after fashion: A Report of AI in Fashion}

\author{%
  Xingxing Zou, Waikeung Wong\thanks{Corresponding author}\\
  The Hong Kong Polytechnic University\\
  Laboratory for Artificial Intelligence in Design (AiDLab)\\
  Hong Kong Science Park, New Territories, Hong Kong \\
  \texttt{xingxingzou@aidlab.hk, calvin.wong@polyu.edu.hk} \\
}

\begin{document}
\maketitle

\begin{abstract}
  In this independent report - \textit{fAshIon} after fashion, we examine the development of \textit{fAshIon} (artificial intelligence (AI) in fashion) and explore its potentiality to become a major disruptor of the fashion industry in the near future.
  To do this, we investigate AI technologies used in the fashion industry – through several lenses.
  We summarise \textit{fAshIon} studies conducted over the past decade and categorise them into seven groups: Overview, Evaluation, Basic Tech, Selling, Styling, Design, and Buying.
  The datasets mentioned in \textit{fAshIon} research have been consolidated on one GitHub page for ease of use\footnote[1]{https://github.com/AemikaChow/DATASOURCE}.
  We analyse the authors' backgrounds and the geographic regions treated in these studies to determine the landscape of \textit{fAshIon} research.
  The results of our analysis are presented with an aim to provide researchers with a holistic view of research in \textit{fAshIon}.
  As part of our primary research, we also review a wide range of cases of applied \textit{fAshIon} in the fashion industry and analyse their impact on the industry, markets and individuals.
  We also identify the challenges presented by \textit{fAshIon} and suggest that these may form the basis for future research.
  We finally exhibit that many potential opportunities exist for the use of AI in fashion which can transform the fashion industry embedded with AI technologies and boost profits.
\end{abstract}

\section*{In Brief}
\label{intro}
According to \textit{AI in Fashion Market Research Report 2021}~\citep{fashionreport}, under the cumulative impact of COVID-19, global spending on AI in the fashion market is expected to grow from USD 229 million in 2019 to USD 1,260 million by 2024, at a Compound Annual Growth Rate (CAGR) of 40.8\% during the forecast period.
Global expenditure on AI in the fashion market is expected to grow from USD 352.58 million in 2020 to USD 825.19 million by the end of 2025.

\begin{figure}
  \centering
  \includegraphics[height=7.35cm]{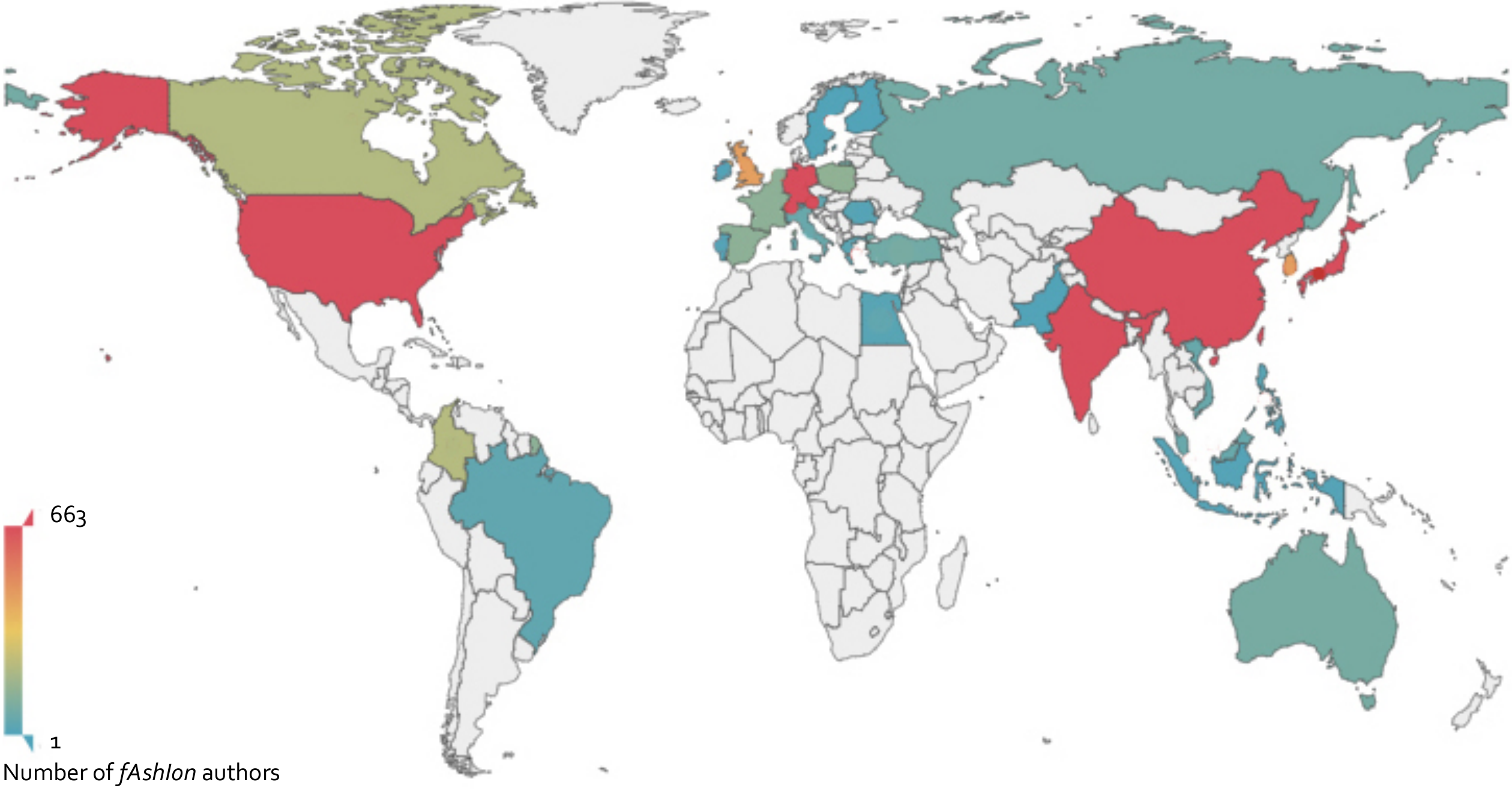}
  \caption{Global heatmap of \textit{fAshIon} researchers' regions based on the 521 papers investigated in this report. The data used to create this figure are derived from a tally of each author's region of affiliation. We find that researchers came from 38 regions.}
  \label{fig:worldmap}
\end{figure}

\textbf{In this report, we analyse the \textit{fAshIon} research and their applications, explore the potential of \textit{fAshIon} to transform the fashion industry and discuss the challenges and opportunities.}
Firstly, we investigate \textit{fAshIon} research papers reported over the past decade.
We search for papers focused on the applications of AI technologies in fashion involving the databases such as IEEE Xplore Digital Library\footnote{https://ieeexplore.ieee.org/Xplore/cookiedetectresponse.jsp}, Google Scholar\footnote{https://scholar.google.com/}, and Arxiv\footnote{https://arxiv.org/}.
The earliest paper reviewed in this report was published on 31 March 2008~\citep{han2008bottom}; the most recent paper was published on 9 March 2021.
Five hundred twenty-one papers are analysed, covering 1,465 authors affiliated with 383 different entities (different departments of the same university or company are not regarded as different affiliations; for example, Amazon Lab126, Amazon Visual Search and AR Group are all counted as Amazon).
As shown in Figure~\ref{fig:worldmap}, China, the United States of America, Singapore, India, Japan, and Germany are identified as the hottest regions of \textit{fAshIon} research.
We present region information here to emphasise that fashion is extremely complicated which may be treated in different ways in different regions. Taking an example of the attribute datasets collected for fashion recognition, the data distribution is greatly different. `Paisley', a specific type of `print', which is a general attribute in the fashion datasets collected from India but may be a long-tailed attribute in the datasets collected from other regions.
In the following, more detailed region information will be presented.
This region information provides the interested parties in \textit{fAshion} with the hidden clues of the application popularity in different regions and the potential opportunities.

Unlike the general `Review Paper', the methods and techniques of related fashion tasks using AI have not been presented in detail one by one.
We organise \textit{fAshIon} studies according to their applications in the fashion industry to clearly show the gap between existing research and practical requirements.
Meanwhile, we point out the hottest topics according to the word cloud based on the titles of papers.
Statistical results of these papers, including the authors, the author's affiliations and their corresponding regions, the years of publication, most popular papers, most active researchers, etc., are presented to provide rich information of \textit{fAshIon} for researchers.
By presenting the analysis of \textit{fAshIon} research, we hope to provide the information about 1. the gap between existing research and practical requirements;  and 2. the hottest topics of each practical direction. As a result, researchers who have interests in this filed can quick start their \textit{fAshIon} research by paying attention on the most active authors, reading the most porpular papers, and utlizing the summary of published fashion datasets\footnote{https://github.com/AemikaChow/DATASOURCE}.
Details can be found in Section~\ref{related}.

Additionally, we also investigate related products and services provided by 126 companies and start-ups.
A wide range of cases of applied \textit{fAshIon} in the fashion industry are reviewed and analysed in terms of their impact on the industry, markets, and individuals.
By presenting the analysis of \textit{fAshIon} applications, we hope to present the following key information: 1. existence of current types of \textit{fAshIon} applications; 2. many technologies not being converted to practical application; 3. existence of barriers between technical frameworks and business models.
Details can be found in Section~\ref{method}.
Finally, we draw conclusions about the challenges and the opportunities that exist in \textit{fAshIon}, which may form the basis for future research.
This detailed analysis is described in Section~\ref{cha} and in Section~\ref{is}.

\section{\textit{fAshIon} Research}
\label{related}
\begin{figure}
  \centering
  \includegraphics[height=4.05 cm]{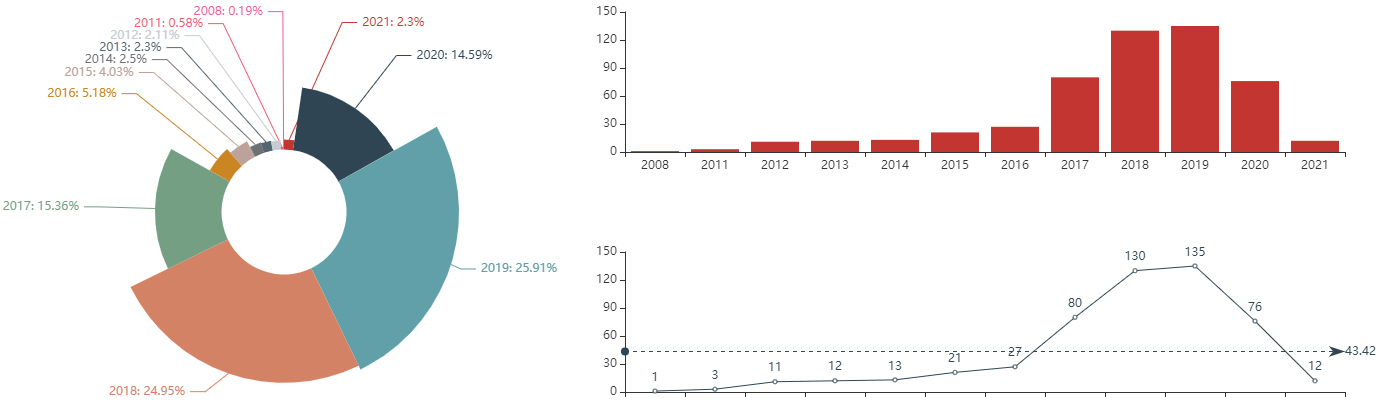}
  \caption{Distribution of published \textit{fAshIon} papers by year. A search identified 521 papers published in the field. Apart from the 12 papers published in 2021 (which is not a final count) and the 76 papers published in 2020, we can see that the overall trend of these papers is growing.}
  \label{fig:year}
\end{figure}

To our knowledge, 521 papers related to \textit{fAshIon} have been published in the last 13 years.
Figure~\ref{fig:year} indicates that the overall trend of \textit{fAshIon}-related papers is one of growth.
The first identified paper on \textit{fAshIon} paper was published in 2008.
Although there is no record of \textit{fAshIon} papers published in 2009 or 2010, the number of papers has been increasing continuously since 2011.
The year 2017 included a particularly conspicuous step in the growth of \textit{fAshIon} research, which is consistent with the conclusion of the fashion market in \citep{fashionreport}.
During the years 2017, 2018, and 2019, 80, 130, and 135 papers on this topic were published, respectively.
The number of papers published in 2020 is lower compared with those in 2019 and 2018 and almost equal to the number published in 2017.
One possible reason for this is the influence of the COVID-19 pandemic, which swept the world in 2020.

\begin{figure}
  \centering
  \includegraphics[height=5.5cm]{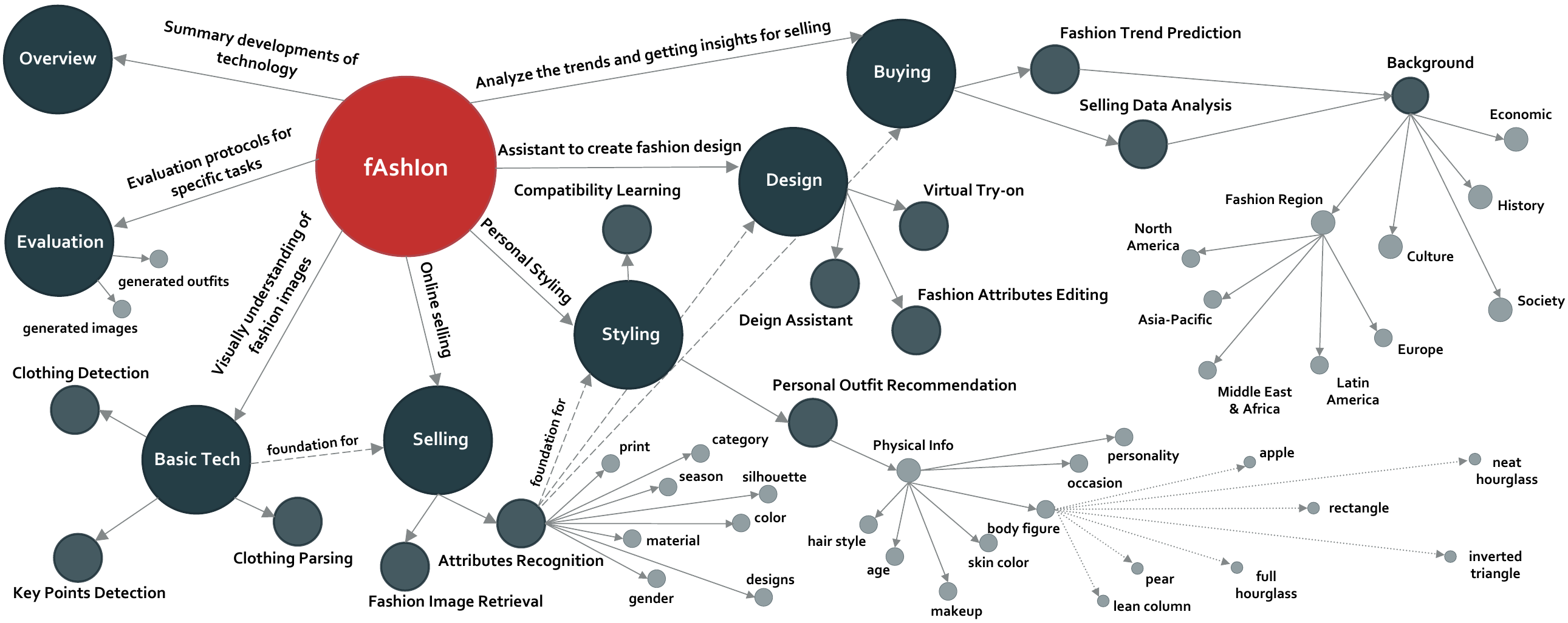}
  \caption{Mind map of \textit{fAshIon} research (some nodes are omitted in this graph for clarity).}
  \label{fig:know}
\end{figure}

Additionally, the gap between these research results and the market’s expectations also may have directly dampened the researchers’ enthusiasm to a certain extent.
After carefully reading about these 521 papers, as shown in Figure~\ref{fig:know}, we categorised them into \textbf{7}  groups, namely: \textbf{Overview, Evaluation, Basic Tech, Selling, Styling, Design, and Buying.}
This taxonomy was inspired by \citep{zou2018fashion}, which divides \textit{fAshIon} research into groups based on existing roles in the fashion industry, namely the fashion lover, the stylist, and the designer.
We organise these papers in this taxonomy to better illustrate the relationship between computer vision problems and fashion tasks.
As shown in Figure~\ref{fig:know}, fashion tasks are extremely complicated since they would be affected by many factors.
For example, `compatibility learning'~\citep{han2017learning} which may help to reduce the workload of a stylist by generating the outfit composition automatically.
Fitting a model to an outfit dataset is not difficult.
The challenge is how to explain the generated results.
Fashion is subjective.
In most situations, a stylist needs a reasonable explanation to persuade customers to believe something or change their mind when they hold a different opinion.
If no explanation to underpin the decision such as evaluation or recommendation, the decision is not convincing enough for the customers to accept.
Naturally, when tackling the compatibility learning task, how to give a concrete reason for the results needs to be taken into consideration.
All in all, as the papers are categorised in this way, it is easier for identifying corresponding roles in the fashion industry and convenient for comparing existing research with practical requirements.

\begin{figure}
  \centering
  \includegraphics[height=4.1cm]{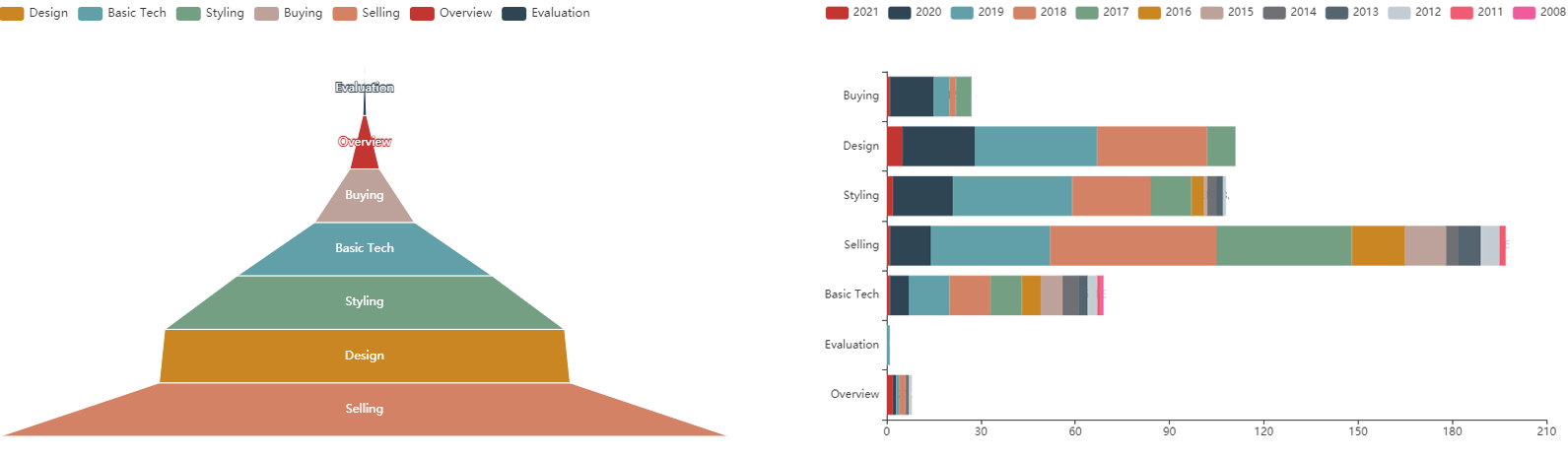}
  \caption{Distribution of published \textit{fAshIon} papers by 7 groups, namely, Overview, Evaluation, Basic Tech, Selling, Styling, Design, and Buying.}
  \label{fig:group}
\end{figure}

Based on this taxonomy, we further analyse the 521 papers.
The distribution of published \textit{fAshIon} papers by these 7 groups is shown in Figure~\ref{fig:group}.
It can be found that most papers were focused on solving problems in the group of `Selling'.
Based on the research interest, these groups sorted from high to low are `Selling', `Design', `Styling', `Basic Tech', `Buying', `Overview', and `Evaluation'.
Furthermore, as shown on the right side of Figure~\ref{fig:group}, we can find more interesting information of \textit{fAshIon} research.
Papers in the group of `Overview' were relatively evenly distributed by year.
Very few researchers have focused on defining the evaluation standard for fashion tasks.
Although solving the problems in the group of `Basic Tech' is not so popular comparing with the other groups  \textit{fAshIon} research, it has still received attention every year.
Additionally, we analyse the group of `Selling', `Styling', `Design', and `Buying' together since all these four groups are closely related to corresponding roles in the fashion industry.
Researchers focused on solving the problems in `Selling' at the earliest (in 2011) and gradually turned their attention to the research on `Styling' (in 2012), `Design' (in 2017), and `Buying' (in 2017).
`Selling' is the most popular topic among these four groups before the year 2018.
The sum of papers in these three groups, i.e. `Styling', `Design', and `Buying', is 27 (i.e., 13 + 9 + 5 = 27) in 2017.
Meanwhile, 53 papers in total related to the tasks in the group of `Selling' were published in 2017.
In other words, in 2017, papers published in the `Selling' group was over the total number of the `Styling', `Design', and `Buying' group, i.e., 43 versus 27.
In the following year 2018, the compared numbers are changed to 53 versus 62 (i.e. 25 + 35 + 2 = 62).
Even though the number of papers under `Selling' group still accounted for the largest proportion of published papers among these four groups in 2018, the situation has started to change in 2019.
The numbers of published papers in the `Selling', `Styling', and `Design' group were very close and almost equal in 2019, which is 38, 38, and 39, respectively. In other words, researchers turned their attention from solving problems in the `Selling' group into tackling tasks related to `Styling' and `Design'.
`Design' became the most popular topics among these four groups after 2019.
In addition, it is worth mentioning that, although research in `Buying' is relatively limited, it is important to the fashion industry as the same as `Selling', `Styling', and `Design'.

On the basis of the analysis above, we conclude that  1. the overall trend of the \textit{fAshIon} papers is growing; 2. research on proposing the evaluation protocol for specific fashion tasks has not yet raised attention; 3. research on solving problems in the group of `Selling' was popular before and slowly lost attention; 4. research on tackling tasks in the group of `Styling' and `Design' becomes increasingly hot in recent years; 5. research on problems related to `Buying' was overall increased year by year. Papers in the group of `Buying' are still relatively few which is not in directly proportional to its importance to the fashion industry.
In the following sections, we present our detailed analysis according to each defined group.
The definition of each group is given first, followed by a word cloud based on the titles of the papers in the group as appropriate.
Then, we present the statistical analysis of papers organised by year of publication, authorial affiliations, the corresponding regions, and the authors.
Although we do not introduce the methods proposed in these papers, the most active authors and the most popular papers (according to the number of citations) are listed as a reference for researchers.

\subsection{Overview}
\begin{figure}
  \centering
  \includegraphics[height=4.4 cm]{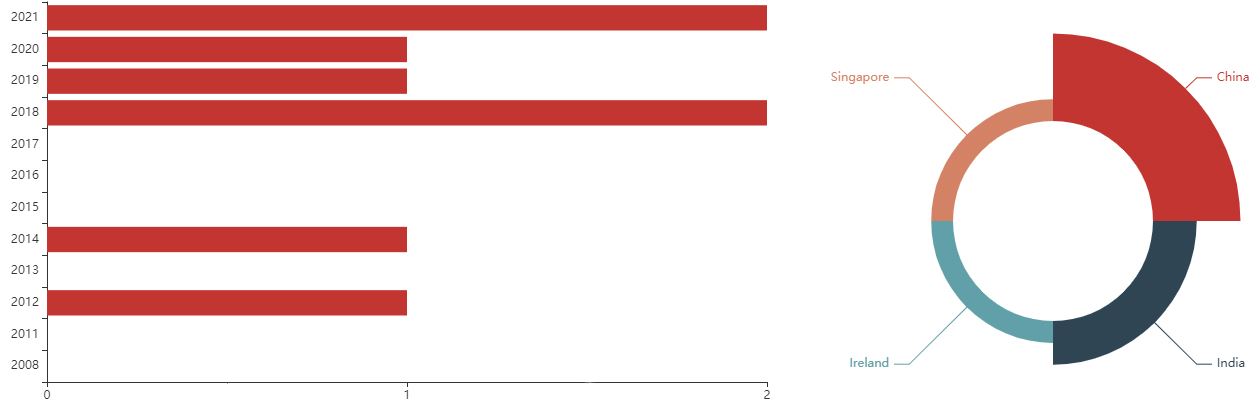}
  \caption{Distribution of Overview papers. These include 4 papers from China, 2 from India, 1 from Ireland, and 1 from Singapore. The number reflects the region of the first author's affiliation rather than the nationality of the first author.}
  \label{fig:overview}
\end{figure}

The papers in this group are review papers~\citep{gong2021aesthetics,kashilani2018overview,liu2014fashion,cheng2020fashion,song2018multimedia} that summarise the developments in fAshIon technology in different areas.
Overview papers can help researchers to quickly understand the current situation of \textit{fAshIon} research and familiarise them with the corresponding methods, related datasets, baseline approaches, and evaluation protocols.
For example, \citep{kashilani2018overview} presents a summary of the various retrieval techniques used in clothing retrieval, and \citep{cheng2020fashion} provides an analysis of the four main tasks in fashion: fashion detection, including landmark detection, fashion parsing, and item retrieval; fashion analysis, consisting of attribute recognition, style learning, and popularity prediction; fashion synthesis, which involves style transfer, pose transformation, and physical simulation; and fashion recommendation, which comprises fashion compatibility, outfit matching, and hairstyle suggestion.
\citep{cheng2020fashion} also presents the benchmark datasets and the evaluation protocols of each task.
\citep{gong2021aesthetics} summarises the use of a recommendation system in situations using different methods such as End-to-End, Implicit Feedback-based, Weak Appearance Feature-based, Semantic Attribute Region-guided, and Using Adversarial Feature Transformer; then, the paper presents some approaches related to understanding aesthetics and realising personalisation in \textit{fAshIon}.
The ratio of Overview papers to the total number of published \textit{fAshIon} papers is around 1.5\%.
As shown in Figure~\ref{fig:overview}, 4 papers from China, 2 from India, 1 from Ireland, and 1 from Singapore are identified.
These detailed numbers are derived from the regions of the first author's affiliation rather than the nationality of the first author; for example, if the first author is affiliated with Amazon Lab126, the paper is counted as being from the United States.
Although not takes a large proportion of the \textit{fAshIon} research,  it still plays a role for researchers to quickly have a holistic view of \textit{fAshIon} from the pespective of technology.

\subsection{Evaluation}
Evaluation protocols are important for research.
Unlike the studies of art, which are extremely subjective, science is quantifiable.
From a research perspective, science undoubtedly needs objective indicators to demonstrate the advantage of one approach to accomplish a specific task.
It is unfair to evaluate a paper as better than another based on subjective criteria.
For example, 'accuracy' is one of the criteria used to evaluate whether a recognition model is good.
Demonstrating the advantage of one model based solely on the retrieved results is not possible.
However, there is a huge contradiction here.
As has been mentioned, fashion is closer to art than science.
The consideration of how to design an evaluation protocol for \textit{fAshIon} research with a high degree of subjectivity has become challenging.
Very few research have specifically investigated the evaluation protocols in \textit{fAshIon} research.
The most mainstream evaluation indicators have been adopted directly from corresponding computer science tasks, such as accuracy for fashion recognition (image recognition tasks) or inception score~\citep{che2016mode} for fashion generation (image generation tasks).
To our knowledge, \citep{sherman2019assessing} is the only paper that describes the process of conducting research from the perspective of evaluation protocol. (Note: the use of `only' does not mean that no other \textit{fAshIon} evaluation indicators have been introduced in other papers.
Fill-in-the-blank accuracy and compatibility prediction AUC~\citep{VasilevaECCV18FasionCompatibility, cucurull2019context, wang2019outfit} are two evaluation indicators that have been introduced for the assessment of fashion compatibility.
The papers in which they are introduced have adopted unique \textit{fAshIon} indicators, but this is not the main goal of the research.) \citep{sherman2019assessing}, published in 2019 by True Fit\footnote{https://www.truefit.com/en/Home} in the USA, is mainly focused on solving the problem of evaluating a fashion recommendation system by including multiple metrics that are relevant to fashion and performing within segments of users with different interaction histories.

\subsection{Basic Tech}
\begin{figure}
  \centering
  \includegraphics[height=7.2cm]{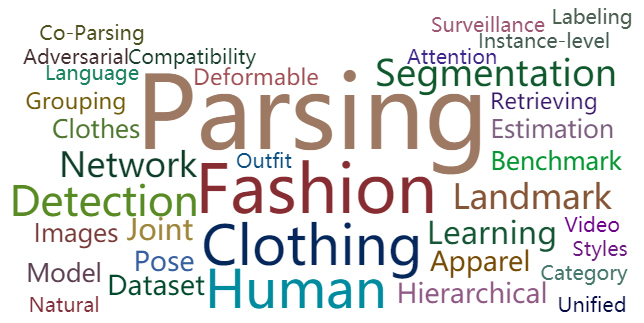}
  \caption{Word cloud based on the titles of papers in the Basic Tech group. Words such as `the', `a', and `toward' were manually deleted. Some words with similar meanings, e.g., `clothing' and `clothings' or `model' and `models', were grouped together.}
  \label{fig:basicword}
\end{figure}

The third \textit{fAshIon} research group is Basic Tech.
Papers in this group are mainly focused on the basic technology used to process fashion images~\citep{tian2021improving, lin2020aggregation, ziegler2020fashion, castro2020segmentation, lee2019global, ge2019deepfashion2, jia2019fashionpedia, yu2019layout, li2019spatial, liang2018look, li2017holistic, xia2017joint, tangseng2017looking, liu2017surveillance, yan2017unconstrained, liang2016clothes, wu2016enhanced, liu2016fashion, dong2015parsing, liu2015fashion, yamaguchi2014retrieving, liu2013fashion, yamaguchi2013paper, yamaguchi2012parsing, wang2011blocks}.
It is also a foundation of all computer vision tasks.
To provide a snapshot of the research in this group, we present a word cloud based on these research titles.
Words such as `the', `a', and `toward' were manually deleted, while some words with similar meanings, e.g., `clothing' and `clothings' or `model' and `models', were grouped together.
It can be seen in Figure~\ref{fig:basicword} that the keywords in this group are Parsing, Fashion, Clothing, Human, Segmentation, and Landmark.
We conclude that the research in Basic Tech focus on image-level processing, e.g., clothing parsing, landmark detection, key point, and apparel detection. The ratio of Basic Tech papers to all \textit{fAshIon} papers is around 13.2\%.

\begin{figure}
  \centering
  \includegraphics[height=4.5cm]{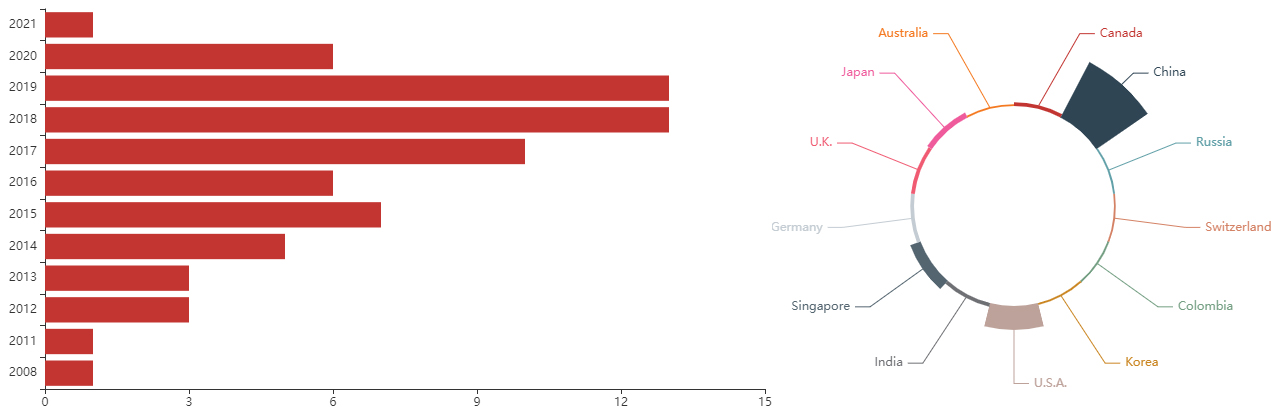}
  \caption{ Distribution of Basic Tech papers. These include 34 papers from China, 13 from the USA, 6 from Singapore, 3 from Japan, 2 from Canada, 2 from India,  2 from Germany, 2 from the UK, 1 from Russia, 1 from Switzerland, 1 from Colombia, 1 from Korea, and 1 from Australia.}
  \label{fig:basic}
\end{figure}

As shown in Figure~\ref{fig:basic}, this group comprises 69 papers, including 34 from China, 13 from the USA, and 6 from Singapore.
The hottest topic in this group is clothing parsing with regard to clothing segmentation in semantic-level~\citep{liang2016clothes,liu2017surveillance,li2017holistic,liang2018look, dong2015parsing,liu2015fashion, yamaguchi2013paper, liu2013fashion, yamaguchi2012parsing} \textit{etc}.
We list the authors who have published papers in this group, sorted from high to low. Here, we emphasise the top 4 authors (taking only the first author into consideration): Xiaodan Liang\footnote{https://lemondan.github.io/} from the School of Intelligent Systems Engineering, Sun Yat-sen University; Si Liu\footnote{https://sites.google.com/site/siliuhome/} from the Institute of Information Engineering, Chinese Academy of Sciences; Jian Dong\footnote{https://sites.google.com/site/homejiandong/} from the Department of Electrical and Computer Engineering, National University of Singapore; and Kota Yamaguchi\footnote{https://sites.google.com/view/kyamagu} from CyberAgent, Inc., previously from Tohoku University.

In Table~\ref{sample-basic}, aside from the numbers of papers published by these authors in this group, we also present the top 3 related papers according to the number of citations.
‘\citep{liang2016semantic}: 236 | 2016' refers to the paper ‘Semantic Object Parsing with Graph LSTM’, which was published in 2016 and has been cited 236 times so far.
‘\citep{yamaguchi2012parsing}: 468 | 2012' refers to the paper ‘Parsing Clothing in Fashion Photographs’, which was published in 2012 and has been cited 468 times.
‘\citep{yamaguchi2013paper}: 272 | 2013' refers to the paper ‘Paper Doll Parsing: Retrieving Similar Styles to Parse Clothing Items’, which was published in 2013 and has been cited 272 times.

Many other authors have also contributed excellent works to this field.
Overall, 227 authors worldwide have been identified; among these, 52 authors have been identified.
The top 12 listed authors with their number of publications (in bracket) are Shicheng Yan (13), Liang Lin (12), Xiaodan Liang (11), Xiaohui Shen (8), Si Liu (7), Jiashi Feng (6), Ping Luo (5), Jianchao Yang (5), Ke Gong (4), Jian Dong (4), Kota Yamaguchi (4), and M. Hadi Kiapour (4).
These works form the foundation of \textit{fAshIon} research, especially in the application of functions to complicated images such as crowd images or images from social media.

These first three groups, Overview, Evaluation, and Basic Tech, are more general and not so closely related to roles in the fashion industry.
Next, we will introduce the tasks involved with similar duties of certain roles in fashion.

\begin{table}
  \caption{List of the top 4 authors who have published papers in the Basic Tech group, sorted from high to low and the most relevant papers according to their citations.}
  \label{sample-basic}
  \centering
  \begin{tabular}{lcccc}
    \toprule
    Researchers     & No. of papers  & (citations | year)  & (citations | year)  & (citations | year)  \\
    \midrule
    Xiaodan Liang   & 6  & \citep{liang2016semantic}: 236 | 2016 & \citep{liang2015human}: 210 | 2015  & \citep{liang2015deep}: 191 | 2015  \\
    Si Liu  & 5  & \citep{liu2013fashion}: 168 | 2013  & \citep{liu2015fashion}: 59 | 2015 & \citep{liu2011weakly}: 50 | 2011  \\
    Jian Dong  & 3  & \citep{dong2013deformable}: 84 | 2013  & \citep{dong2014towards}: 81 | 2014 & \citep{dong2015parsing}: 9 | 2015 \\
    Kota Yamaguchi   &  3 & \citep{yamaguchi2012parsing}: 468 | 2012  & \citep{yamaguchi2013paper}: 272 | 2013  & \citep{yamaguchi2014retrieving}: 124 | 2014\\
    \bottomrule
  \end{tabular}
\end{table}

\subsection{Selling}
In this section, we introduce the tasks involved with similar duties of certain areas in fashion, namely Selling, Styling, Design, and Buying.
The seller takes a main role in the fashion retail sector to sell products.
A good seller in fashion needs to recommend accurate products to his or her customers.
Their aim is to increase sales as much as possible.
The basic requirements for the seller are therefore:

\begin{itemize}
  \item That he or she can recognise the detailed attributes of fashion products and know how to find similar items according to a general description using simple words provided by his or her customers.
  \item That he or she can accurately recommend fashion products according to observations on both his or her customers and of trends.
  \item That he or she can reply quickly to his or her customers about whether a particular item or size is available or out of stock.
\end{itemize}

\begin{figure}
  \centering
  \includegraphics[height=7.5cm]{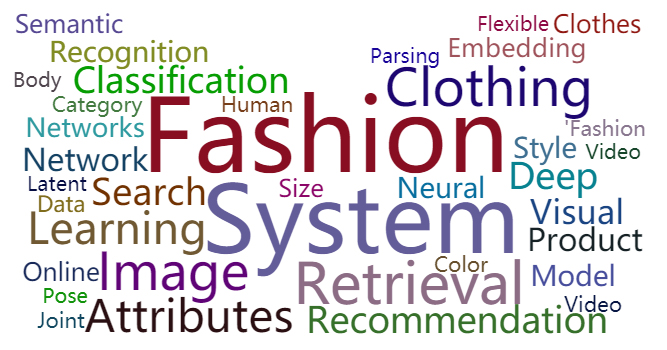}
  \caption{Word cloud based on the titles of papers in the Selling group. The keywords are Fashion, System, Clothing, Retrieval, Image, Attributes, and Recommendation.}
  \label{fig:sellerword}
\end{figure}

According to these requirements, we identify many research papers that aim to provide methods to create online virtual sellers.
These papers are then categorised into the Selling group~\citep{bossard2012apparel, di2013style, miura2013snapper, jagadeesh2014large, finegrained, loni2014fashion, kiapour2014hipster, SimoSerraCVPR2015, hadi2015buy, huang2015cross, xiao2015learning, lin2015rapid, yamaguchi2015mix, vittayakorn2016automatic, liuLQWTcvpr16DeepFashion, liu2016mvc, StreetStyle2017, takagi2017makes, han2017automatic, InoueICCVW2017, laenen2017cross, gu2017understanding, cheng2017video2shop, gu2018multi, manandhar2018tiered, hadi2018brand, kuang2018ontology, zou2019fashionai, guo2019imaterialist, chen2020tailorgan}.
It can be seen in Figure~\ref{fig:sellerword} that the keywords of this group are Fashion, Clothing, System, Image, Attributes, Retrieval, Recommendation, and Classification.
We conclude that the research in Selling has been focused on marketing fashion products like a seller, e.g., by using fashion item recognition, recommendation, and retrieval.
The ratio of Selling papers to the total number of papers is around 37.8 \%; accounting for the largest proportion of research among all of the seven groups identified in the current papers.

\begin{figure}
  \centering
  \includegraphics[height=4.5cm]{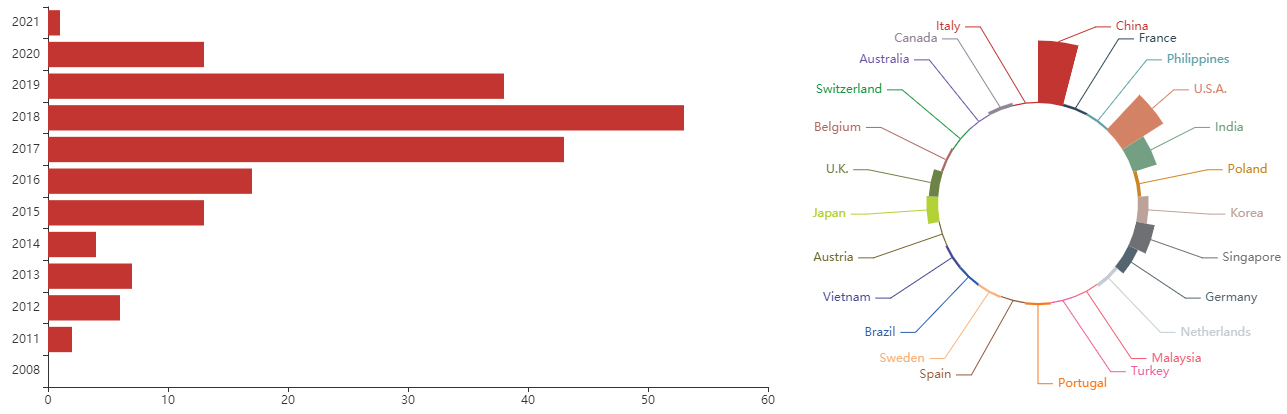}
  \caption{Distribution of Selling papers. This group includes 53 papers from China, 41 from the USA, 21 from India, 16 from Singapore, 10 from Japan, 9 from Korea, 9 from Germany, 9 from the UK, 3 from Poland, 3 from Canada, 3 from the Netherlands, 2 from France, 2 from the Philippines, 2 from Portugal, 2 from Sweden, 2 from Brazil, 2 from Vietnam, 2 from Belgium, 1 from Malaysia, 1 from Turkey, 1 from Spain, 1 from Austria, 1 from Switzerland, 1 from Australia, and 1 from Italy.}
  \label{fig:seller}
\end{figure}

This group includes 197 papers.
As shown in Figure~\ref{fig:seller}, these include 53 papers from China, 41 from the USA, 21 from India, 16 from Singapore, 10 from Japan, 9 from Korea, 9 from Germany, and 9 from the UK. China, the USA, and India are therefore identified as the hottest regions for this type of research.
Large bodies of research were published in 2017, 2018, and 2019.
Attribute recognition and image retrieval are identified as the hottest topics in this group.

We list the authors who have published papers in this group, sorted from high to low. Here, we emphasise the top 3 authors (taking only the first author into consideration): Huijing Zhan\footnote{https://zhanhuijing.github.io/} from DL2.0 of the Institute for Infocomm Research (I2R); M. Hadi Kiapour\footnote{http://www.cs.unc.edu/~hadi/} from Dawnlight, previously from eBay Inc.; and Qi Dong\footnote{https://www.aminer.cn/profile/qi-dong/53f42d48dabfaeb1a7b89950} from the Recognition and Video team at AWS, Amazon (previously from Queen Mary University of London).

In Table~\ref{sample-seller}, in addition to the numbers of papers published by the authors in this group, we also present the top 3 related papers according to the number of citations.
Particularly, we emphasise some papers that have received much attention.
`Image-based Recommendations on Styles and Substitutes~\citep{mcauley2015image}' was published in 2015 and has been cited 1,089 times.
`DeepFashion: Powering Robust Clothes Recognition and Retrieval with Rich Annotations~\citep{liuLQWTcvpr16DeepFashion}' was published in 2016 and has been cited 857 times.
`Describing Clothing by Semantic Attributes~\citep{chen2012describing}' was published in 2012 and has been cited 412 times.
`Street-to-shop: Cross-scenario Clothing Retrieval via Parts Alignment and Auxiliary Set~\citep{liu2012street}' was published in 2012 and has been cited 370 times.
`Where to Buy it: Matching Street Clothing Photos in Online Shops~\citep{hadi2015buy}' was published in 2015 and has been cited 361 times.
`Whittlesearch: Image search with Relative Attribute Feedback~\citep{kovashka2012whittlesearch}' was published in 2012 and has been cited 336 times.
`Fine-grained visual comparisons with local learning~\citep{yu2014fine}' was published in 2014 and has been cited 304 times.

Many other authors have also contributed excellent works to this field.
Overall, 574 authors are identified worldwide, including 151 first authors.
The top 16 listed authors are Huijing Zhan (11), M. Hadi Kiapour (3), Qi Dong (3), Shuohao Li (2), Sanyi Zhang (2), Dipu Manandhar (2), Ruifan Li (2), Zhi-Qi Cheng (2), Kenan E. Ak (2), Roshanank Zakizadeh (2), Julia Lasserre (2), Shuhui Jiang (2), Kaori Abe (2), Xiaoling Gu (2), Sirion Vittayakon (2), and Junshi Huang (2).
Their works have focused on the most general fashion scenarios with regard to selling products.

This body of work aims to provide a better shopping experience to online shoppers, thus improving the performance of sales.
In addition, this kind of research provides a foundation for higher-level tasks.
The ability to execute fine-grained fashion attributes recognition comprises the basic knowledge needed to deal with higher-level tasks such as mixing and matching.

\begin{table}
  \caption{List of the top 3 authors who have published papers in the Selling group, sorted from high to low and the most relevant papers according to their citations.}
  \label{sample-seller}
  \centering
  \begin{tabular}{lcccc}
    \toprule
    Researchers     & No. of papers  & (citations | year)  & (citations | year)  & (citations | year)  \\
    \midrule
    Huijing Zhan   & 11  & \citep{zhan2019deepshoe}: 10 | 2019 & \citep{zhan2017cross}: 6 | 2017  & \citep{zhan2017street}: 4 | 2017  \\
    M. Hadi Kiapour  & 3  & \citep{hadi2015buy}: 365 | 2015  & \citep{kiapour2014hipster}: 219 | 2014 & \citep{hadi2018brand}: 3 | 2018  \\
    Qi Dong  & 3  & \citep{dong2018imbalanced}: 134 | 2018  & \citep{dong2017class}: 83 | 2017 & \citep{dong2017multi}: 60 | 2017 \\
    \bottomrule
  \end{tabular}
\end{table}

\subsection{Styling}
Generally speaking, a stylist who provides styling advice to customers should have good beauty sense and an ability to provide personal styling services to individual customers.
They should be able to recognise fashion attributes.
The basic requirements for a stylist are:

\begin{itemize}
  \item That he or she has a good sense of clothing aesthetics, such as a sense of colour, a sense of texture, and a sense of silhouette, and can easily create a well-composed outfit. (Understand Clothing Aesthetic)
  \item That he or she knows how to make an outfit attain visual balance for a given customer and mix and match according to different situations, such as for an occasion, the seasons, or body figure. (Do Personal Styling)
  \item That he or she understands what beauty is and can convincingly explain his or her selections, enabling him or her to persuade the customer and provide satisfactory service. (Good Story Teller)
  \item That he or she always stays aware of fashion trends and can apply these trends to his or her work.
\end{itemize}

\begin{figure}
  \centering
  \includegraphics[height=7.5cm]{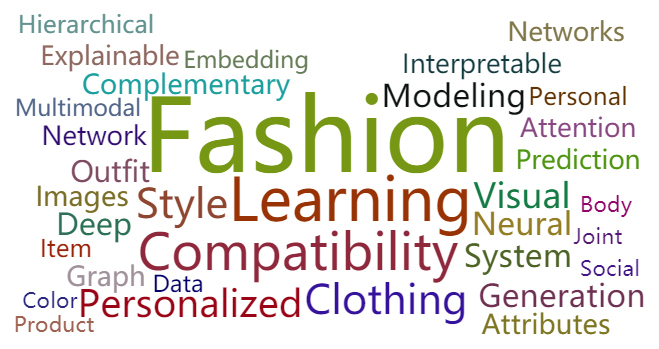}
  \caption{Word cloud based on the titles of papers in the Styling group. The keywords are Fashion, Learning, Compatibility, Clothing, and Personalized.}
  \label{fig:stylistword}
\end{figure}

According to these requirements, we identify many research papers that aim to provide online styling services.
We then categorise these papers into the Styling group~\citep{hazra2020attr2style,verma2020addressing,hidayati2020dress,al2020paris,sachdeva2020interactive,liu2020learning,zhang2020learning,kim2020self,sagar2020pai,lin2020fashion,li2020hierarchical,zheng2020personalized,tangseng2020toward,singhal2020towards,laenen2020attention,stefani2019cfrs,li2019coherent,kang2019complete,cui2019dressing,bettaney2019fashion,lin2020fashion,polania2019learning,sun2020learning,lin2019learning,wang2019outfit,sonie2019personalised,chen2019pog,zhang2018clothes,han2017learning,zou2020regularizing}.
It can be seen in Figure~\ref{fig:stylistword} that the keywords of this group are Fashion, Clothing, System, Image, Attributes, Retrieval, Recommendation, and Classification. We conclude that the research in Styling has been focused on providing online styling services, e.g., by using fashion compatibility learning, outfit creation, and recommendation. The ratio of Styling papers to the total number of papers is around 20.7\%.

\begin{figure}
  \centering
  \includegraphics[height=4.5cm]{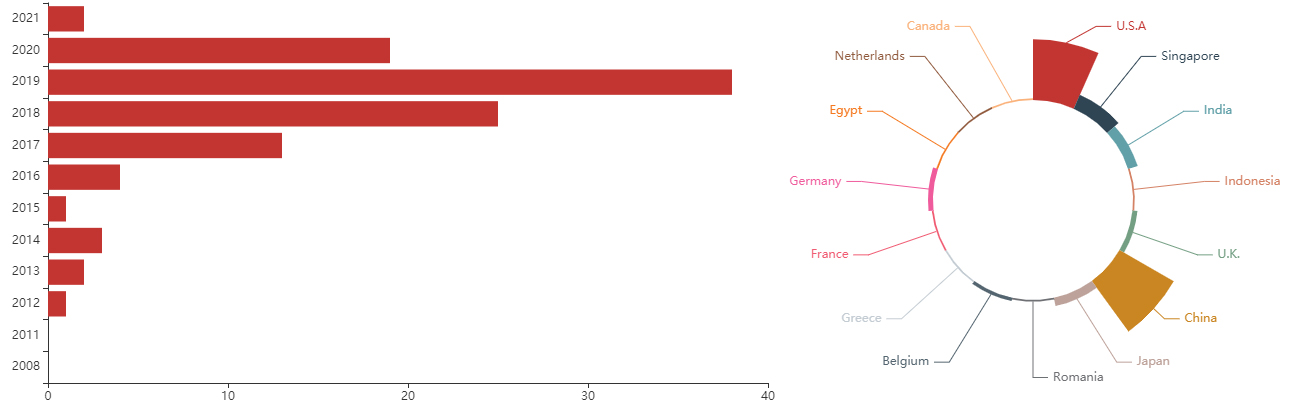}
  \caption{Distribution of Styling papers. These include 37 papers from China, 36 from the USA, 9 from Singapore, 6 from India, 5 from Japan, 3 from the UK, 3 from Germany, 2 from Belgium, 1 from Indonesia, 1 from Romania, 1 from Greece, 1 from France, 1 from Egypt, 1 from the Netherlands, and 1 from Canada.}
  \label{fig:stylist}
\end{figure}

As shown in Figure~\ref{fig:stylist}, this group comprises 108 papers, including 37 papers from China, 36 from the USA, 9 from Singapore, 6 from India, and 5 from Japan. China, the USA, and Singapore are identified as the hottest regions for this type of research. Large bodies of research were published in 2017, 2018, and 2019.
Compatibility learning and outfit recommendation are identified as the hottest topics in this group.

We also list the authors who have published papers in this group, sorted from high to low. Here, we emphasise the top 5 authors (taking only the first author into consideration): Xuemeng Song\footnote{https://xuemengsong.github.io/} from Shandong University, Xun Yang\footnote{https://www.researchgate.net/scientific-contributions/Xun-Yang-2151358935} from the National University of Singapore, Wang-Cheng Kang\footnote{http://cseweb.ucsd.edu/~wckang/} from Google Research (Brain team), Ruining He\footnote{https://sites.google.com/view/ruining-he/} from Google Research, and Pongsate Tangseng\footnote{http://vision.is.tohoku.ac.jp/~tangseng/} from the Graduate School of Information Sciences, Tohoku University.

In Table~\ref{sample-stylist}, aside from the number of papers published by these authors in this group, we also present the top 3 related papers according to the number of citations.
Particularly, we emphasise a paper which has received much attention: `Ups and downs: Modeling the Visual Evolution of Fashion Trends with One-Class Collaborative Filtering~\citep{he2016ups}' was published in 2016 and has been cited 894 times.
`Hi, Magic Closet, Tell me What to Wear!~\citep{liu2012hi}' was published in 2012 and has been cited 272 times.
`Neuroaesthetics in Fashion: Modeling the Perception of Fashionability~\citep{SimoSerraCVPR2015}' was published in 2015 and has been cited 196 times.
`Learning Fashion Compatibility with Bidirectional LSTMs~\citep{han2017learning}' was published in 2017 and has been cited 175 times.
`Large Scale Visual Recommendations from Street Fashion Images~\citep{jagadeesh2014large}' was published in 2014 and has been cited 119 times.

Many other authors have also contributed excellent works to this field.
Overall, this group includes 454 authors worldwide, including 120 first authors.
The top 9 authors in this group are Julian McAuley (7), Tat-Seng Chua (5), Xun Yang (4), Yunshan Ma (4), Xuemeng Song (4), Xingnan He (4), Reza Shirvany (4), Urs Bergmann (4), and Jun Ma (4).
Works in this group focus on duties, such as those of a stylist in the fashion industry.

The works aim to provide personalised online styling services that can cross-sell to increase the exposure rate of products (outfit recommendation) and to create new business models that offer either an online mix-and-match service or complete outfit boxes based on the personal preference of the customer (do personal styling).

\begin{table}
  \caption{List of the top 3 authors who have published papers in the Styling group, sorted from high to low and the most relevant papers according to their citations.}
  \label{sample-stylist}
  \centering
  \begin{tabular}{lcccc}
    \toprule
    Researchers     & No. of papers  & (citations | year)  & (citations | year)  & (citations | year)  \\
    \midrule
    Xuemeng Song   & 6  & \citep{song2017neurostylist}: 99 | 2017 & \citep{song2018neural}: 70 | 2018  & \citep{song2019gp}: 7 | 2019   \\
    Xun Yang  & 3  & \citep{yang2019interpretable}: 22 | 2019  & \citep{ma2019and}: 13 | 2019 & \citep{yang2020learning}: - | 2020  \\
    Wang-Cheng Kang  & 2  & \citep{kang2017visually}: 113 | 2017  & \citep{kang2019complete}: 25 | 2019 & -  \\
    Ruining He  & 2  & \citep{he2016ups}: 894 | 2016  & \citep{he2016learning}: 60 | 2016 & - \\
    Pongsate Tangseng   & 2  & \citep{tangseng2017recommending}: 29 | 2017 & \citep{tangseng2020toward}: 6 | 2020  & -  \\
    \bottomrule
  \end{tabular}
\end{table}

\subsection{Design}
In the fashion industry, a good design depends on the abilities of a fashion designer. Generally speaking, the designer is most closely related to the stylist and needs to have similar aesthetic abilities to those of the stylist. More importantly, a designer should know not only what beauty is but also how to create it.
The unique requirements for a designer are:

\begin{itemize}
  \item That he or she has the ability to transform highly abstract concepts or stories into fashion items (e.g., garments, accessories, bags, shoes, etc.) through the language of clothing.
  \item That he or she has the ability to transfer images or feelings from his or her brain to papers via his or her hand drawing or computer software.
  \item That he or she has the ability to create new designs based on a theme or conditioned by different constraints, such as the seasons, the trends, or the DNA of the brand  (Brand DNA is the essence of your identity as a business).
  \item That he or she always remains aware of fashion trends and can apply these trends to his or her work.
\end{itemize}

\begin{figure}
  \centering
  \includegraphics[height=7.5cm]{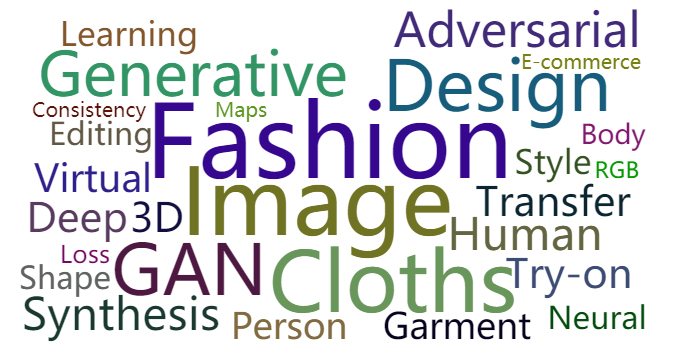}
  \caption{Word cloud based on the titles of papers in the Design group. The keywords are Fashion, GAN, Cloths, Design, Image, Generative, Adversarial, Try-On, and Synthesis.}
  \label{fig:designerword}
\end{figure}

According to these requirements, we identify many research papers that aim to help fashion designers, which we categorise into the Design group~\citep{yoon2021neural,wu2021example,kuppa2020shineon,su2020deepcloth,li2020deep,li2020deep1,caliskan2020multi,tango2020anime,dubey2020ai,huang2020arch,jiang2020bcnet,dong2020fashion,zhan2020pose,han2020design,yuan2020garment,zhang2020deep,patel2020tailornet,chen2020tailorgan,yuan2020garment,gundogdu2020garnet++,hsiao2019fashion++,dong2019fashion,kato2019gans,gu2019ladn,santesteban2019learning,shi2019learning,yu2019personalized,pandey2020poly,natsume2019siclope,liu2019swapgan,zhou2019text,dong2019towards,liu2019toward,zhao2018compensation,li2018beautygan,hu2018deep,siarohin2018deformable,sbai2018design,cui2018fashiongan,chou2018pivtons,xian2018texturegan,han2018viton,zhu2017your,jiang2017fashion,date2017fashioning,lorbert2017toward,han2019finet,han2019clothflow,dong2018soft}.
It can be seen in Figure~\ref{fig:designerword} that the keywords of this group are Fashion, GAN, Cloths, Design, Image, Generative, Adversarial, Try-On, and Synthesis.
We conclude that the research in Design focus on fashion generation, virtual try-on, clothing synthesis.
The ratio of Design papers to the total number of papers is around 21.3\%.

\begin{figure}
  \centering
  \includegraphics[height=4.4cm]{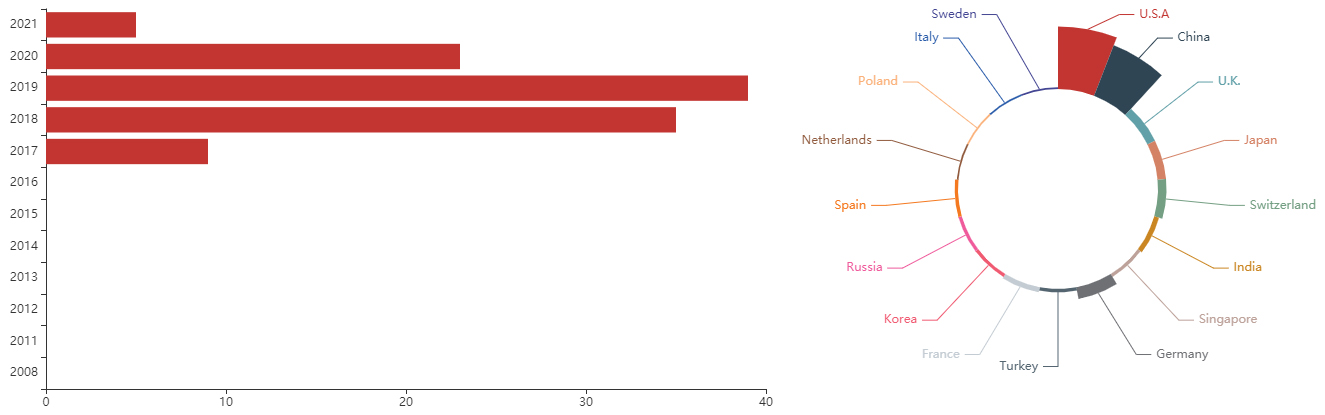}
  \caption{Distribution of Design papers. These include 37 papers from the USA, 32 from China, 7 from Germany, 5 from the UK, 5 from Japan, 5 from Switzerland, 3 from France, 3 from India, 2 from Singapore, 2 from Turkey, 2 from Korea, 2 from Russia, 2 from Spain, 1 from the Netherlands, 1 from Poland, 1 from Italy, and 1 from Sweden.}
  \label{fig:designer}
\end{figure}

As shown in Figure~\ref{fig:designer}, this group comprises 111 papers, including 37 from the USA, 32 from China, 7 from Germany, 5 from the UK, 5 from Japan, and 5 from Switzerland. The USA, China, and Germany are identified as the hottest regions for this type of research. Large bodies of research were published in 2018, 2019, and 2020. Virtual Try-On (VITON), Neural Style Transfer, Image Synthesis, and 3D Clothing Generation are identified as the hottest topics in this group.
Moreover, we list the authors who have published papers in this group, sorted from high to low. Here, we emphasise the top 4 authors (taking only the first author into consideration): Haoye Dong\footnote{https://www.scholat.com/donghaoye} from Sun Yat-sen University; Xintong Han\footnote{http://users.umiacs.umd.edu/~xintong/} from Huya Inc., previously from the University of Maryland; Erhan Gundogdu\footnote{https://egundogdu.github.io/} from Amazon Berlin; and Natsumi Kat\footnote{https://digitalnature.slis.tsukuba.ac.jp/2017/04/natsumi-kato/} from the University of Tsukuba.

In Table~\ref{sample-designer}, aside from the number of papers published by these top authors in the group, we also present the top 3 related papers according to the number of citations.
Particularly, we emphasise a paper that has received much attention: ‘Disentangled Person Image Generation~\citep{ma2018disentangled}’ was published in 2018 and has been cited 267 times.
`Deformable GANs for Pose-based Human Image Generation~\citep{siarohin2018deformable}' was published in 2018 and has been cited 211 times.
`A Variational U-Net for Conditional Appearance and Shape Generation~\citep{esser2018variational}' has published in 2018 and has been cited 207 times.
`VITON: An Image-based Virtual Try-on Network~\citep{han2018viton}' was published in 2018 and has been cited 177 times.
`BodyNet: Volumetric Inference of 3D Human Body Shapes~\citep{varol2018bodynet}' was published in 2018 and has been cited 174 times.
`PIFU: Pixel-Aligned Implicit Function for High-Resolution Clothed Human Digitization~\citep{saito2019pifu}' was published in 2019 and has been cited 172 times.
`Be Your Own Prada: Fashion Synthesis with Strucral Coherence~\citep{zhu2017your}' was published in 2017 and has been cited 166 times.

Many other authors have also contributed excellent works to this field.
Overall, this group includes 427 authors worldwide, of which 99 are first authors.
The top 10 listed authors are Xiaodan Liang (7), Haoye Dong (7), Gerard Pons-Moll (5), Jian Yin (4), Zeng Huang (3), Xintong Han (3), Duygu Ceylan (3), Tony Tung (3), Hao Li (3), and Christian Theobalt (3).
These works have focused on helping the fashion designer, particularly via image generation, style transfer, and 3D garment generation.

Most of current works focus on helping designers or online retailers to visually demonstrate their ideas, e.g., attribute editing~\citep{dong2020fashion,su2020deepcloth,hsiao2019fashion++}, 3D garment generation~\citep{patel2020tailornet,caliskan2020multi,yoon2021neural,varol2018bodynet,gundogdu2019garnet,gundogdu2020garnet++}, virtual try-on~\citep{santesteban2019learning,dong2019fw,kuppa2020shineon,dong2019towards,chou2018pivtons,han2018viton}, etc.
These methods can speed up the process of design in terms of time saving on design detail modification to a certain extent.
Furthermore, material waste can be reduced since making up a physical sample for each design is not necessary.
In addition, marketing materials such as videos, brochures, flyers, etc. for promoting fashion products of retailers can be generated by using GAN models and thus marketing costs involved in shooting, promotional material design, etc. can be reduced.
Certainly, with a holistic view of the `Design' research, there are many interesting but challenging tasks remain.
Can AI models create new designs based on a theme or subject to different predetermined constraints, such as seasons, trends, brand images, etc.?
Can AI models transform highly abstract concepts or themes into fashion items, such as garments, accessories, bags, shoes, etc.?

\begin{table}
  \caption{List of the top 3 authors who have published papers in the Design group, sorted from high to low and the most relevant papers according to their citations.}
  \label{sample-designer}
  \centering
  \begin{tabular}{lcccc}
    \toprule
    Researchers     & No. of papers  & (citations | year)  & (citations | year)  & (citations | year)  \\
    \midrule
    Haoye Dong  & 7 & \citep{dong2018soft}: 75 | 2018 & \citep{dong2019towards}: 50 | 2019  & \citep{hu2018deep}: 47 | 2018   \\
    Xintong Han  & 3  & \citep{han2018viton}: 177 | 2018  & \citep{han2019finet}: 27 | 2019 & \citep{han2019clothflow}: 27 | 2019  \\
    Erhan Gundogdu  & 2  & \citep{gundogdu2019garnet}: 39 | 2019  & \citep{gundogdu2020garnet++}: 1 | 2020 & - \\
    Natsumi Kato   & 2  & \citep{kato2018deepwear}: 12 | 2018 & \citep{kato2019gans}: 6 | 2019  & -  \\
    \bottomrule
  \end{tabular}
\end{table}

\subsection{Buying}
The last group is Buying. In the fashion industry, a buyer performs the action of buying the right products for fashion brands and retailers in the retail sector.
Unlike a stylist or a designer, a buyer’s work depends on making the decision to buy the right products in the right quantities.
The unique requirements for a buyer are:

\begin{itemize}
  \item That he or she can draw conclusions from the records of previous sales.
  \item That he or she has great insight into the market and the ability to make predictions.
  \item That he or she is very experienced in fashion retail and fully knows the market trends.
\end{itemize}

\begin{figure}
  \centering
  \includegraphics[height=6.2 cm]{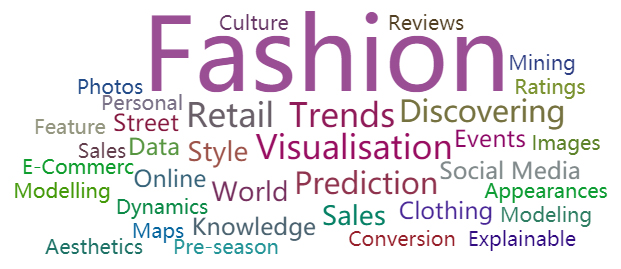}
  \caption{Word cloud based on the titles of papers in the Buying group. The keywords are Fashion, Trends, Discovering, Visualisation, Prediction, and Social Media.}
  \label{fig:buyerword}
\end{figure}

According to these requirements, we identify many research papers on the subject of Buying~\citep{hsiao2021culture,mall2020discovering,al2020modeling,takahashi2020cat,sajja2021explainable,ma2020knowledge,vashishtha2020product,sung2020breaking,singh2019fashion}.
It can be seen in Figure~\ref{fig:buyerword} that the keywords of this group are Fashion, Trends, Discovering, Visualisation, Prediction, and Social Media.
We conclude that the research in Buying focus on trend prediction and culture discovering.
The ratio of Design papers to the total number of papers is around 5.2\%.

\begin{figure}
  \centering
  \includegraphics[height=4.5cm]{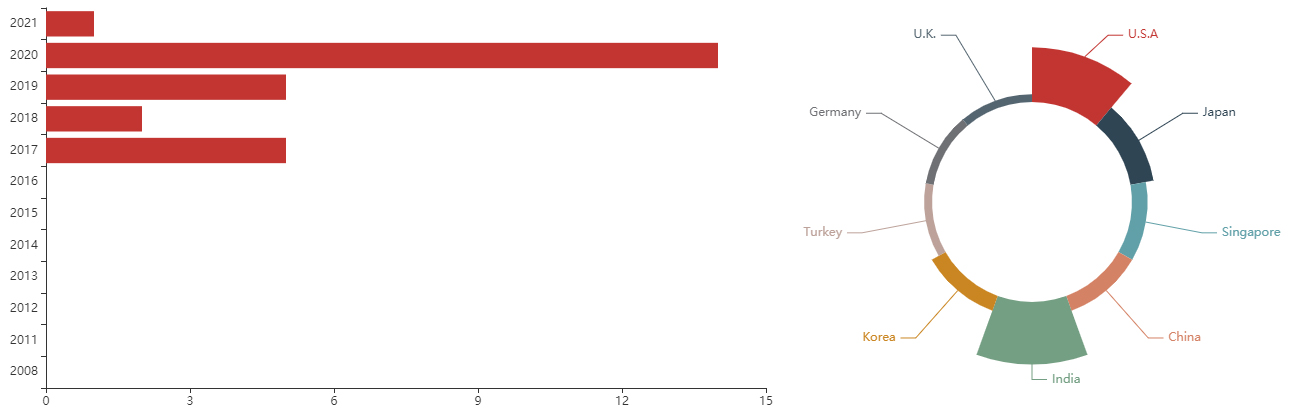}
  \caption{Distribution of Buying papers. These include 8 papers from India, 7 from the USA, 3 from Japan, 2 from Singapore, 2 from Korea, 2 from China, 1 from Turkey, 1 from the UK, and 1 from Germany.}
  \label{fig:buyer}
\end{figure}

As shown in Figure~\ref{fig:buyer}, 27 papers are categorised in this group, including 8 papers from India, 7 from the USA, and 3 from Japan.
India, USA, and Japan are identified as the hottest regions for this group.
A large body of studies was published in 2020.
Trend forecasting is identified as the hottest topic in this group.

Moreover, we present a list of authors who have published papers in this group, sorted from high to low. This group comprises 81 authors worldwide, including 17 first authors. The top 4 listed authors are Ziad AI-Halah (3), Utkarsh Mall (2), Sajan Kedia (2), and Shin Woong Sung (2).
The research in this group aim to mimic the duties of a fashion buyer.
Utilising the power of data-driven techniques to speed up information filtering and the sorting and analysis of big data can greatly benefit fashion retailers, including commercial fashion brands.
Most research in this area has been focused on fashion forecasting, prediction, and selling.

\section{\textit{fAshIon} Application}
\label{method}
\begin{figure}
  \centering
  \includegraphics[height=7.9 cm]{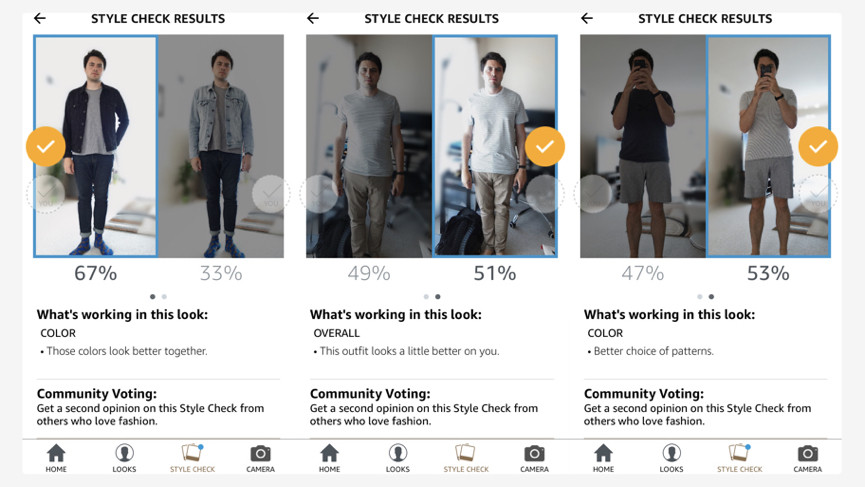}
  \caption{ Echo Look, an outfit evaluation service provided by Amazon (source: Google Images).}
  \label{fig:echo}
\end{figure}

In addition to our research on \textit{fAshIon}, we also focus on related companies and start-ups.
Based on our investigation, 126 companies and start-ups that offer \textit{fAshIon}-related services, applications, or technology support have emerged in recent years.
For example, Echo Look by Amazon (Figure~\ref{fig:echo}) provides customers with a special online service that compares two uploaded outfits and selects the better outfit using AI.
Thus, the user can determine the best look for the day.
The technology used in this application belongs to the previously described Styling group.
Other products and functions, such as Fashion Box provided by STITCH FIX\footnote{https://www.stitchfix.com/}, utilise similar technologies.
After completing a style quiz, the customer receives a box of clothing items recommended for them, providing a different shopping experience from before.
Some companies, such as Alibaba, focus on outfit generation and recommendation, presenting outfit compositions on the checkout page for cross-selling.

In this section, we introduce the opportunities for \textit{fAshIon} based on an analysis of the products or services provided by the investigated companies.
Existing applications or products in the industry are presented briefly.
Detailed information can be found on their official websites.
We do not present the statistical results for these investigated companies because these numbers are largely inconsequential and because the website links of some of the investigated companies are now invalid, such as that of Glitch\footnote{https://glitch-ai.com/}, an AI clothing brand founded by computer scientists turned fashion designers at MIT.
We review a wide range of cases of applied \textit{fAshIon} in the fashion industry and categorized them according to their corresponding technologies (i.e., Selling, Styling, Design, Buying).
We summarise the current situations in \textit{fAshIon} and analyse its impact on the industry, markets and individuals in the end of this section.

\begin{figure}
  \centering
  \includegraphics[height=6.3 cm]{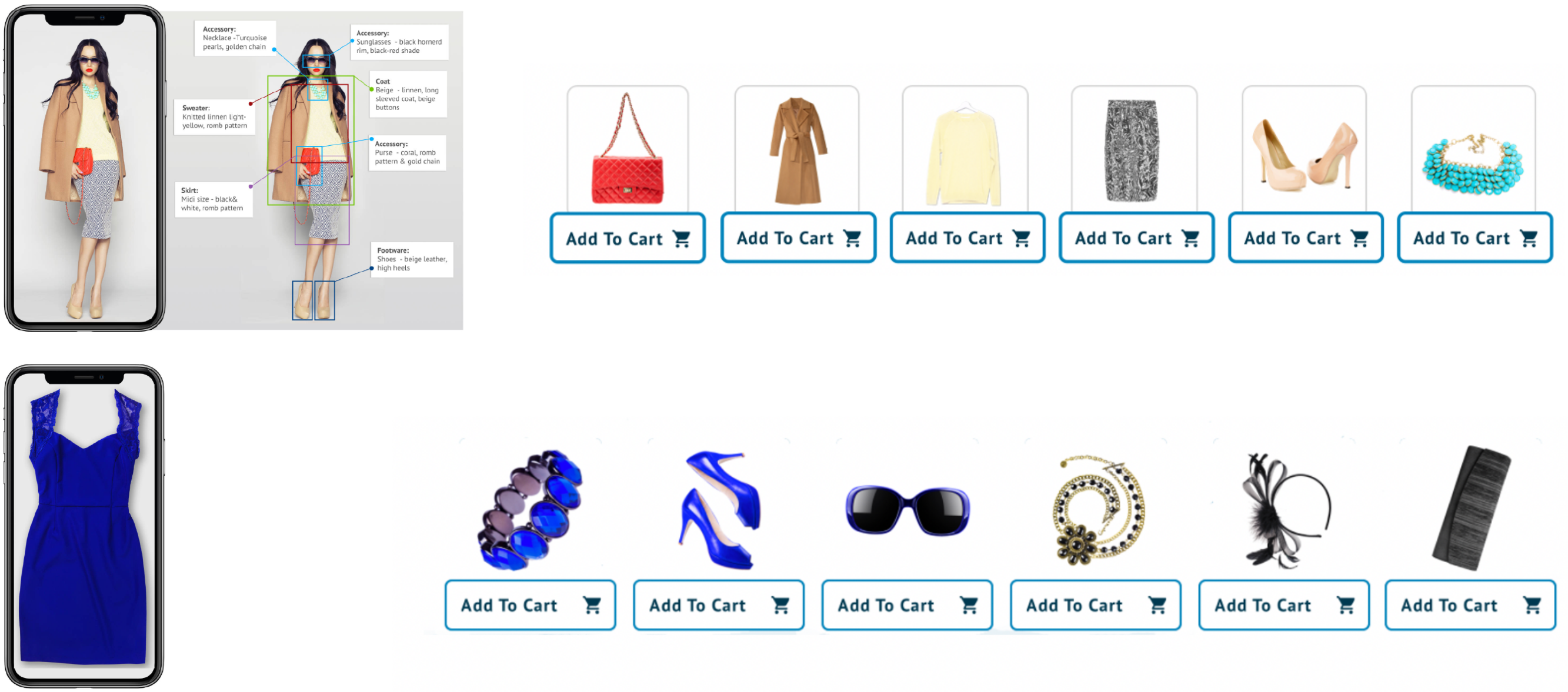}
  \caption{Case study of Fashion Searching and Complete the Look by MACTY. Such platforms enable one-click searches to obtain a full outfit and can also learn customer preferences, promote bundle up and sell up, and improve the searching experience. This technology can help to inspire users who are considering what to wear and how to combine pieces by automatically creating the perfect look.}
  \label{fig:ret}
\end{figure}

\noindent\textbf{Fashion Searching and Complete the Look  - `Selling'}: Technology can help customers to search for fashion items immediately conditioned on uploaded images. Utilising the recognition model to perform attributes tagging can also speed up the process of lunching new products. Additionally, algorithms can make fashionable recommendations for achieving a complete look, a feature that may be widely applicable in cross-selling. A case study of Fashion Searching and Complete the Look by MACTY (Figure~\ref{fig:ret}) reveals that the platform can enable one-click searching to obtain a full outfit. This technology can help to inspire users who are considering what to wear and how to combine pieces by automatically creating the perfect look.

More information can be found at catchoom\footnote{https://catchoom.com/}; chicengine\footnote{http://www.chicengine.com/}; intelistyle\footnote{https://www.intelistyle.com/}; MACTY\footnote{https://macty.eu/index.html}; Markable AI\footnote{https://markable.ai/}; OMINIOUS\footnote{https://www.omnious.com/}; SNAPVISION\footnote{https://www.snap.vision/}; Streamoid\footnote{https://www.streamoid.com/}; syte\footnote{https://www.syte.ai/}; Truefit\footnote{https://www.truefit.com/en/Home}; VISENZE\footnote{https://www.visenze.com/}; WIDE EYES\footnote{https://wideeyes.ai/}; ximilar\footnote{https://www.ximilar.com/}

\begin{figure}
  \centering
  \includegraphics[height=3.6 cm]{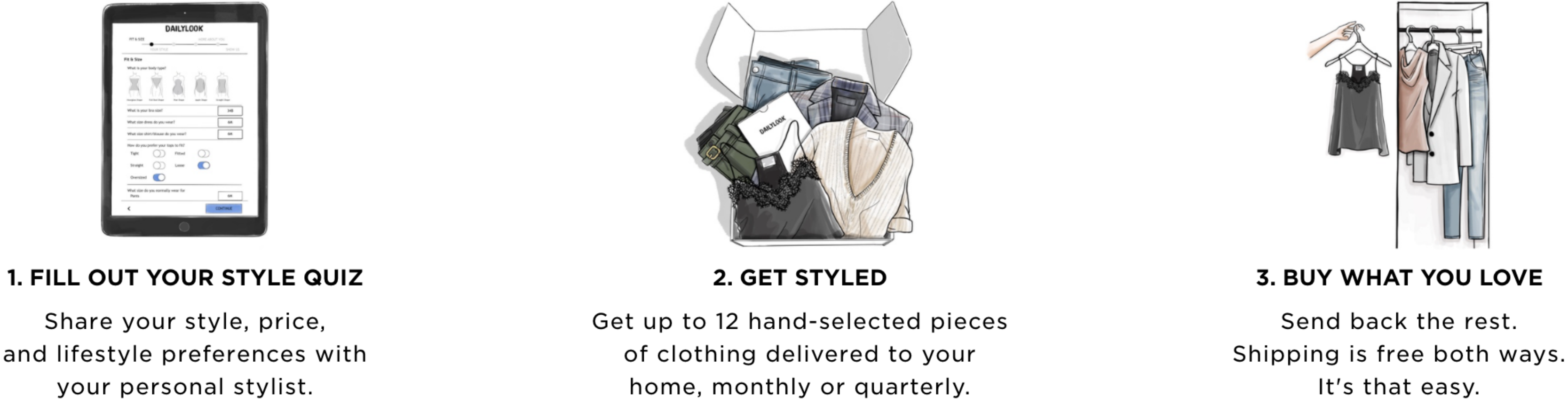}
  \caption{Case Study of a Stylist Service provided by DAILYLOOK. Technology can help users to find their perfect fit and achieve a one-of-a-kind style by selecting from among exclusive brands hand-selected by an expert stylist.}
  \label{fig:out}
\end{figure}

\noindent\textbf{Stylist Service - `Styling'}:
As shown in Figure~\ref{fig:out}, technology can help users to find their perfect fit and achieve one-of-a-kind style by selecting items from the exclusive brands that are hand-selected by an expert stylist.

More information can be found at COUCHFASHION\footnote{https://deck.couchfashion.com/}; STITCH FIX\footnote{https://www.stitchfix.com/}; TRUNK CLUB\footnote{https://www.trunkclub.com/}; Vue.ai\footnote{https://vue.ai/}; DAILYLOOK\footnote{https://www.dailylook.com/}; frockbox\footnote{https://www.frockbox.ca/}; MODABOX\footnote{https://moda-box.com/}

\begin{figure}
  \centering
  \includegraphics[height=3.6 cm]{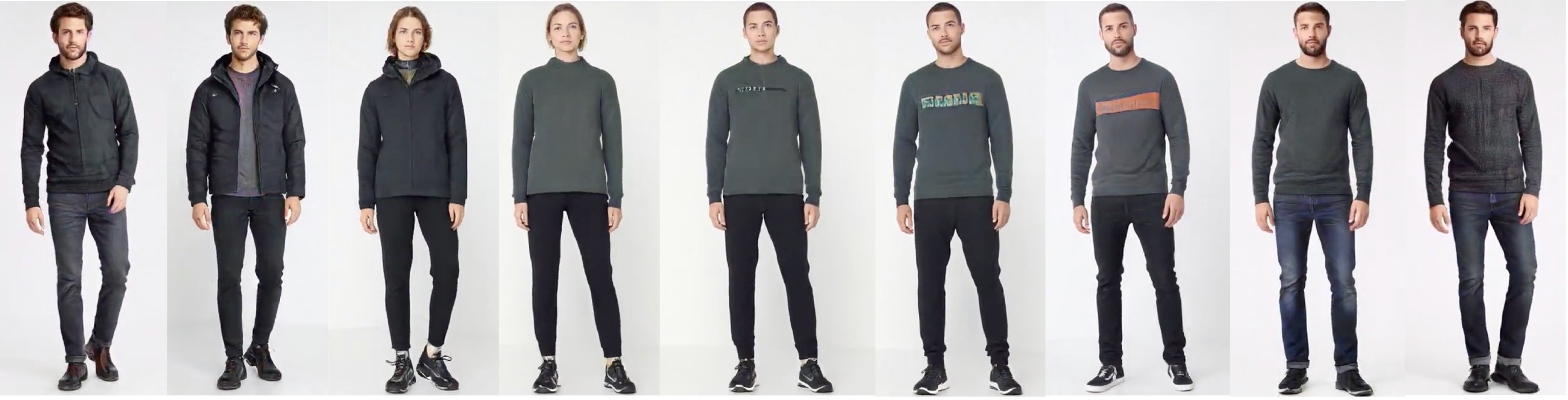}
  \caption{Case study of Virtual Fitting by DATAGRID. Here, AI is used to generate non-existent photorealistic digital humans, and the company aims to utilise it as a new interface for machines in a future society. This type of work can also be used as an advertising model for apparel e-commerce.}
  \label{fig:virtual}
\end{figure}

\noindent\textbf{Virtual Fitting - `Design'}: Technology can improve customers' online shopping experiences by allowing them to virtually visualise the apparel.
Fashion brands and retailers can also use this technology to generate sale or advertising images for their new products without taking new photographs.
Virtual Fitting by DATAGRID (Figure~\ref{fig:virtual}) uses AI to generate non-existent, photorealistic digital humans.
This technology may be utilised as a new interface for machines in a future society where AI has permeated everything. Such work can also be used as an advertising model for apparel e-commerce. Moreover, technology forms an important part of the ‘Virtual Retail' business model, which can accelerate the transformation of the fashion industry from selling Products into selling Service.
Notably, 2D and 3D versions of this technology have now been developed.
Certainly, there are some attempts of utilising neural networks to create virtual fashions which will be transformed into real clothing; e.g., `Project Muse\footnote{https://techcrunch.com/2016/09/02/googles-new-project-muse-proves-machines-arent-that-great-at-fashion-design/}' designed by Google in partnership with Zalando in 2016.

More information can be found at ARMOI\footnote{https://www.armoi.com}; DATAGRID\footnote{https://datagrid.co.jp/en/}; ELSE Corp\footnote{https://www.else-corp.com/}; secret sauce\footnote{https://secretsaucepartners.com/}; 3DLOOK\footnote{https://3dlook.me/}; BIGTHINX\footnote{https://bigthinx.com/}; BOLD METRICS\footnote{https://www.boldmetrics.com/}; TOZI\footnote{https://www.emtailor.com/}; VIRTUSIZE\footnote{http://www.virtusize.com/site/}

\noindent\textbf{Business Strategy - `Buying'}: Technology can help businesses to make predictions by using consumer data from various sources and adjusting businesses' selling strategies to drive revenue while maintaining profitability. Retail buyers, planners, and merchandisers can use this type of technology to react faster to market trends, obtain competitive assortment benchmarking, and optimise prices.
Technology can analyse data to understand the emotional context behind shoppers' purchases; businesses can then deliver the most relevant, personalised experiences to their consumers.
Automatically generating the fashion trending report will be one of the most useful tools.

More information can be found at APPIER\footnote{https://www.appier.com/}; EDITED\footnote{https://edited.com/}; ELEMENTAI\footnote{https://www.elementai.com/}; Liaro\footnote{https://liaro.ai/}; Lilyai\footnote{https://www.lily.ai/}; MANTHAN\footnote{https://www.manthan.com/}; STYLESAGE\footnote{https://stylesage.co/}; stylumia\footnote{https://www.stylumia.ai/}; MAKER/SIGHTS\footnote{https://www.makersights.com/}.

\begin{figure}
  \centering
  \includegraphics[height=7 cm]{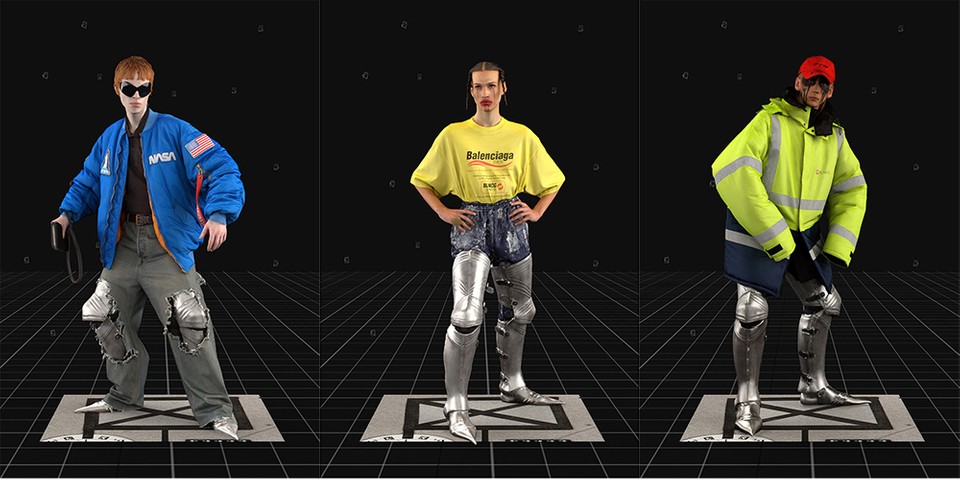}
  \caption{Balenciaga Fall 2021 - a first-of-its kind hybrid video game fashion show. Vogue said: With Balenciaga’s New Gaming App Afterworld Demna Gvasalia Makes the `Quantum Leap' the Industry Has Been Waiting for.}
  \label{fig:ba}
\end{figure}

From the above summary, we can see that AI has already been applied for practical use and some conventional fashion business models have been changed. One of the most typical cases is `fashion searching'.
Based on the technology in the group of `Selling'~\citep{bossard2012apparel, di2013style, miura2013snapper, jagadeesh2014large, loni2014fashion, kiapour2014hipster, SimoSerraCVPR2015, hadi2015buy, huang2015cross, xiao2015learning, lin2015rapid, yamaguchi2015mix, liuLQWTcvpr16DeepFashion, liu2016mvc, takagi2017makes, han2017automatic, laenen2017cross, gu2017understanding, cheng2017video2shop, manandhar2018tiered, hadi2018brand, kuang2018ontology, zou2019fashionai, guo2019imaterialist}, online retailers or online shopping platforms can provide their customers with `intelligent sales person'.
Comparing with the basic requirements for the seller, the AI model can recognise the attributes of fashion products and retrieve the related results.
This technology undoubtedly affects the business model of online selling.
Meanwhile, it is apparent that there is still big room for us to put efforts for improvement, e.g., recommendation of fashion products based on the fashion trends and observations on the customers.
Additionally, many technologies have not been applied; e.g., attributes editing~\citep{dong2020fashion,su2020deepcloth,hsiao2019fashion++} in the group of `Design'.
Especially the COVID-19 has changed the conventional practice of the industry.
During the global pandemic, many large-scale activities were forced to cancel.
Many design houses debuted their new collections via digitalised ways like films or games.
For example, as shown in Figure~\ref{fig:ba}, Belenciaga debuted its fall 2021 collection through the `record-breaking video game,' which is titled, `Afterworld: The Age of Tomorrow\footnote{https://www.vogue.com/fashion-shows/fall-2021-ready-to-wear/balenciaga}.'
\textit{Demna Gvasalia} (creative director of Balenciaga) depicted the world of 2031 in this game. The concept of `parallel world' is utilised to catch a glimpse of the near future based on our understanding about the middle Ages.

\textit{Art at its most significant is a Distant Early Warning System that can always be relied on to tell the old culture what is beginning to happen to it.} \\
\rightline{\textit{-- Marshall McLuhan}} \\

\section{Challenges}\label{cha}
\textit{fAshIon} is a challenging field for computer vision. In addition to the most basic function of clothing, i.e.,  covering the body, well dress-up can bring the feeling of beauty to us.
Fashion is an art form, just like music, painting, poetry, etc. Although creating AI models to assist with the different types of works done by fashion practitioners, including sellers, stylists, designers, and buyers can empower the current industry by upgrading its structure while elevating the economy, to teach an AI model to understand fashion is not an easy thing.
The first question that should be answered is: `What is beauty and how to evaluate it?'
Philosophers have been discussing this issue for millennia.

Different perspectives, such as those of the materialist and the idealist, are juxtaposed and have shaped the meaning of aesthetics.
It is generally recognised that beauty produces a feeling that is similar to happiness.
The beauty or deformity of an object is caused by its gene.
Generally, to perceive beauty, one must perceive the essence that beauty from both internal and external perspectives. Internal meaning differs from external meaning.
The external senses may detect traits that do not rely on any priori perception.
In other words, if you do not perceive or at least grasp the object, its beauty cannot be perceived.
Perception is a major keyword in aesthetics~\cite{reid1850essays}.

\textit{Haute Couture gowns possess the unique individuality of an object d'art.
They are among the last items made by hand, the human hand, whose value is irreplaceable, because it gives its creations that which no machine can ever give: poetry and life.} \\
\rightline{\textit{-- Christian Dior, 1957}} \\

We agree with \textit{Dior}. No matter how far technology advances, the creative capacity of humans is irreplaceable.
In terms of computers, strong AI technology remains a controversial topic, and most mainstream viewpoints oppose conducting related research.
Unlike humans, computers cannot express themselves in an emotional context.
Creating beautiful and valuable artwork is a complicated challenge for a computer because it has no ability to express emotions.
Fortunately, perceiving an object's nature or its structure differs from perceiving its beauty~\cite{shelley2017concept}.
Therefore, if we only focus on the essential concept, i.e., beauty is balance or harmony among the fundamental elements of fashion in an item of apparel or an outfit, \textit{fAshIon} tasks can be completed to a certain extent by a computer.
\textbf{Naturally, the basic knowledge learning serves as the foundation of all higher-order fashion tasks.}

\begin{figure}
  \centering
  \includegraphics[height=4.4 cm]{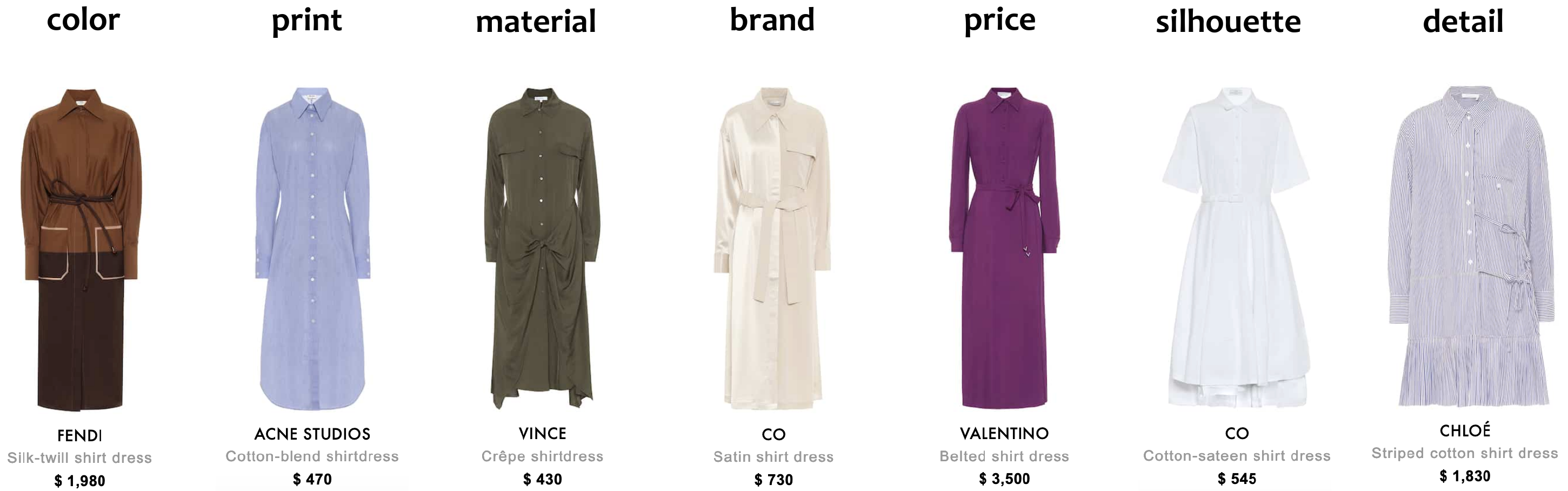}
  \caption{Examples of the many different design elements – e.g., material, print, colour – that are involved in the making of a shirt dress.}
  \label{fig:att}
\end{figure}

However, even though we do not take `perception' into consideration, only visually understanding fashion still remains many challenges.
Specifically, it is difficult to obtain a clear definition of what constitutes the basic knowledge in fashion; e.g., `attributes', `styles', `trends', and `design themes'.
Here, we will present some easily comprehensible examples as a simple demonstration.
The 7 shirt dresses shown in Figure~\ref{fig:att} are obviously very different.
The attributes that are most immediately obvious are the differences in their colours, branding, and prices.
Furthermore, some of the dresses have different prints; the first shirt dress is colour-blocked, whereas the second and the last dresses are striped and the others are solid colours.
Moreover, the materials used to make these shirt dresses are also different: the second dress is made of cotton, and the fourth is made of satin.
Their silhouettes also differ: the fourth dress is H-line and the sixth is A-line.
Thus far, we have only considered the main differences.
If we investigate in further detail, we find that the waists of these 7 dresses are all different from each other.
The design of the pockets and the opening designs are also different.
There are many aspects that make these shirt dresses different from each other.
We have only described a few such aspects here.
If we imagine these design attributes as letters in an alphabet, then the garment is a word made up of these letters.
Then, an outfit is like a sentence composed of many words, all obeying a fixed grammar.
Many outfits gather together to express a certain theme, such as a collection in a fashion show, tell a story composed of many sentences. Now, we ask two questions:

\begin{itemize}
  \item Is the love of Romeo and Juliet related to the 26 letters of the Roman alphabet?
  \item Is Hamlet\footnote{https://en.wikipedia.org/wiki/Hamlet} defined as a tragedy only because of some of the words or sentences used in the play?
\end{itemize}

\begin{figure}
  \centering
  \includegraphics[height=11.2 cm]{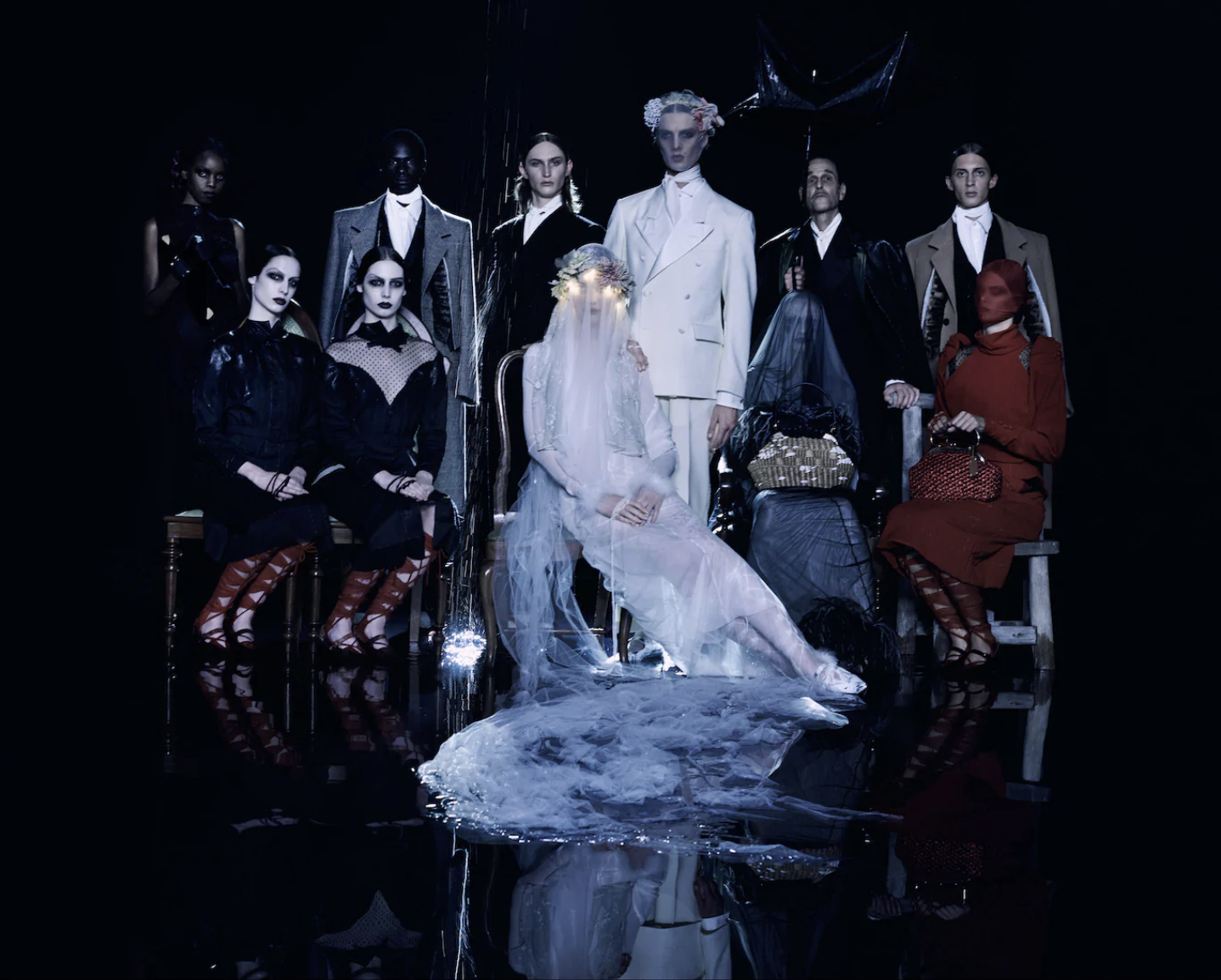}
  \caption{\textbf{S. W. A. L. K. II} (Sealed With A Loving Kiss, Maison Margiela Co-Ed SS21 by John Galliano). Group picture. Standing, from left to right: Look 39, Look 37, Look 35, Look 30, Look 31, Look 33. Sitting, from left to right: Look 36A, Look 36B, Look 29, Look 32, Look 34.}
  \label{fig:mm}
\end{figure}

Here, we use the show \textbf{S. W. A. L. K. II}\footnote{https://www.maisonmargiela.com/us} to explore similar concepts.
The link below presents the related media and information that can be easily find, such as a paragraph describing this show:

\textit{The connectivity of interdependence becomes revalued in times of separation. The reliance of one person upon another is a vital pas de deux activated by instinct and trust. For the Co-Ed Collection Spring-Summer 2021, Maison Margiela interprets this concord through the tango. Vigorous and intense, the dance is cathartic: releasing the spirit of the old, it inspires the lust to move on. It compels acceptance; it heralds new beginnings; it beckons change.}\\

We present some of the looks in this series in Figure~\ref{fig:mm}. How did the designer use tulle, velvet, worn, silk, chiffon, leather, and other materials to express the theme of this show, as mentioned above?

\begin{itemize}
  \item Are the feelings or emotions evoked by this collection closely related to the attributes of fashion that have been adopted?
  \item Can the theme expressed by this fashion show be understood by considering only Look 29 (a white stretch-tulle circular cut dress with farmed plumed trims worn over a white muslin, circular cut long-sleeved dress with white chiffon ghillies and white painted leather Tango pumps) or Look 30 (an ecru wool, double-breasted tuxedo with silk duchesse lapels worn over a white tuxedo shirt with white painted leather Hyperion ankle-strap shoes)?
\end{itemize}

As described in the summary of the hierarchy of visual understanding\footnote{https://informationisbeautiful.net/2010/data-information-knowledge-wisdom/}: the first, or lowest, level in the hierarchy is data (discrete elements), the second level is information (linked elements), the third level is knowledge (organised information), and the highest level is wisdom (applied knowledge). \textbf{The challenges increase progressively at higher levels of visual understanding, and may include:}

\begin{itemize}
  \item Recognising data (visual recognition of fashion images).
  \item Turning data into effective information (obtaining fashion-related information from the recognised attributes, e.g., recognising a style or evaluating an outfit and knowing the behind reason for this evaluation judgement).
  \item Translating information into knowledge based on understanding (understanding clothing aesthetics, knowing how to predict trends and why trends occur).
  \item Applying information freely and forming insights (knowing how to create beauty and lead trends).
\end{itemize}

\section{Opportunities}
\label{is}
\textit{fAshIon} is a fresh field that full of opportunities.
As summarised in the last Section, there has long way to go, even if only visually understand fashion.
Undoubtedly, large amount of problems remain for researchers to tackle.
In additon, under the impact of the epidemic, embracing technology becoming the new trend in fashion.
The whole industry will be unavoidably be reshaped in the Post-COVID-19 Era.
Addition to the Balenciage Fall 2021, many other designers also expressed their own thinkings to the \textit{fAshIon} in the near future; e.g., after Prada 2021 SS\footnote{https://www.vogue.com/fashion-shows/spring-2021-ready-to-wear/prada}, there has `a conversation' in which \textit{Miuccia} talked her ideas about the relationship between humans and machines.
We do not forecast the future of \textit{fAshIon} in the industry due to our limited abilities. Here we would like to use what \textit{Demna Gvasalia} said to represent our viewpoints: \textit{`I believe in a future that is spiritual. Loading a forgotten past.'}
In the following, we conduct a focused analysis of opportunities for research and summarise the possible directions:

\noindent\textbf{Opportunities in `Evaluation':}
`In God we trust; all others must bring data\footnote{W. Edwards Deming}'. As mentioned before, very few research have focused on the evaluation protocols in \textit{fAshIon}.
In addition to the requirements for the algorithms proposed for a specific task to be compared fairly, a company also needs clear criteria to evaluate whether a model is qualified to be embedded in their online products or whether the modelled numbers or predictions can be trusted.

\noindent\textbf{Opportunities in `Basic Tech':}
Even though the research in the group of `Basic Tech' are not directly related to a role in fashion, it is the foundation for computer to understand fashion images.
As mentioned before, most studies targeted on clothing parsing~\citep{liang2016semantic,liang2015deep,liang2015human,dong2015parsing,liu2015fashion,yamaguchi2013paper,dong2014towards,liu2011weakly,yamaguchi2012parsing,wu2016enhanced,liang2016clothes,liu2017surveillance,li2017holistic,liang2018look,dong2020fashion,han2008bottom,liu2013fashion,dong2013deformable}.
However, in terms of clothing parsing, none of them paid attention to the segmentation between the front piece and the back piece of a given single clothing item in a fashion image.
Meanwhile, when several garments are dressed by a person at the same time, it arises a occlusion problem.
In addition, it is great help if the model can segment the garment according to its design regions; e.g., neck region, sleeve region, body region, cuff region, shoulder region, chest region,  waist region etc.

\noindent\textbf{Opportunities in `Selling':} From Figure~\ref{fig:mm}, it is easy to see that there are many attributes of fashion that should be recognised.
This is a multi-label classification task. We roughly estimate that the number of classification labels is over 250~\cite{zou2019fashionai}.
Moreover, there is no existing open-access datasets of comprehensive and well-annotated fashion attributes.
Thus, the efforts to train an AI model using current data will face challenges such as weakly labelled data and noises.
The diverse format of fashion image data, including fashion show images, product images, model images, street photos, and social media images, also increases the difficulty of recognition.
Here we would like to recall the basic requirements for the seller: 1. can recognise the detailed attributes of fashion products and know how to find similar items according to a general description using simple words provided by his or her customers; 2. can accurately recommend fashion products according to observations on both his or her customers and of trends; 3. can reply quickly to his or her customers about whether a particular item or size is available or out of stock.
Therefore, the conclude the possible future directions including: 1. multi-label classification in weakly supervised manner; 2. fine-grained attributes recognition in webly data or videos; 3. image retrieval in massive data.

\noindent\textbf{Opportunities in `Styling':}
Current research in the group of `Styling' are mainly focused on creating outfit composition for online recommendation.
The basic requirements for a stylist are: 1. has a good sense of clothing aesthetics, such as a sense of colour, a sense of texture, and a sense of silhouette, and can easily create a well-composed outfit; 2. knows how to make an outfit attain visual balance for a given customer and mix and match according to different situations, such as for an occasion, the seasons, or body figure; 3. understands what beauty is and can convincingly explain his or her selections, enabling him or her to persuade the customers and provide satisfactory service; 4. always stays aware of fashion trends and can apply these trends to his or her work.
Therefore, we conclude the possible future directions including:
1. evaluating an outfit is good or not with convicing explanation; 2. understanding about the basic `principles' of mix and match and behind logic; 3. recommending personalised styling advice and catching up the trend in time.

\noindent\textbf{Opportunities in `Design':}
Most of current works focus on helping designers or online retailers to visually demonstrate their ideas, e.g., attribute editing~\citep{dong2020fashion,su2020deepcloth,hsiao2019fashion++}, 3D garment generation~\citep{patel2020tailornet,caliskan2020multi,yoon2021neural,varol2018bodynet,gundogdu2019garnet,gundogdu2020garnet++}, virtual try-on~\citep{santesteban2019learning,dong2019fw,kuppa2020shineon,dong2019towards,chou2018pivtons,han2018viton}, etc.
Certainly, with a holistic view of the `Design' research, there remain many interesting but challenging tasks.
Can AI models create new designs based on a theme with different predetermined constraints, such as seasons, trends, brand image, etc.?
Thus, we conclude the possible future directions including:
1. utilising generation methods in virtual marketing; 2. appling generation methods in the process of fashion design; 3. expressing a concept or idea using fashion language.

\noindent\textbf{Opportunities in `Buying':}
Most research in this area focus on fashion forecasting, prediction, and selling~\citep{hsiao2021culture,mall2020discovering,al2020modeling,takahashi2020cat,sajja2021explainable,ma2020knowledge,vashishtha2020product,sung2020breaking,singh2019fashion}.
As a good buyer,  he or she should: 1. can draw conclusions from records of previous sales; 2. have great insight into the market and the ability to make predictions; 3. very experienced in fashion retail and fully knows the market trends.
Therefore, we conclude the possible future directions including:
1. predicting the market trends based on the analysis of massive data; 2. understanding the culture and inferring the effect of an event on fashion which can lead the trend of fashion.

`Big data can tell us what is wrong, not what is right\footnote{N. Taleb}'.
Therefore, we should be very sceptical of any `big data analyst' or `data scientist' who claims to be able to explain a system in a particular domain without the requisite domain expertise or intimate knowledge of the underlying system under consideration\footnote{https://medium.com/@adambreckler/in-god-we-trust-all-others-bring-data-96784d01e9be}.
To really help the industry, it is better to design tasks and solutions based on the domain knowledge.

\begin{ack}
  This research was funded by the Laboratory for Artificial Intelligence in Design (Project Code: RP3-1), Hong Kong Special Administrative Region. The title, \textit{fAshIon after fashion}, was inspired by a seminar given by Professor Hazel Clark at Parsons School of Design to accompany the excellent exhibition fashion after Fashion. The authors would also like to express deep appreciation to Professor Zowie Broach at the Royal College of Art for her enlightening comments and inspirational contributions from the very beginning of the journey of \textit{fAshIon}.
\end{ack}
\newpage
\small
\bibliographystyle{plainnat.bst}
\bibliography{egbib}

\begin{thebibliography}{189}
\providecommand{\natexlab}[1]{#1}
\providecommand{\url}[1]{\texttt{#1}}
\expandafter\ifx\csname urlstyle\endcsname\relax
  \providecommand{\doi}[1]{doi: #1}\else
  \providecommand{\doi}{doi: \begingroup \urlstyle{rm}\Url}\fi

\bibitem[Al-Halah and Grauman(2020{\natexlab{a}})]{al2020modeling}
Ziad Al-Halah and Kristen Grauman.
\newblock Modeling fashion influence from photos.
\newblock \emph{IEEE Transactions on Multimedia}, 2020{\natexlab{a}}.

\bibitem[Al-Halah and Grauman(2020{\natexlab{b}})]{al2020paris}
Ziad Al-Halah and Kristen Grauman.
\newblock From paris to berlin: Discovering fashion style influences around the
  world.
\newblock In \emph{Proceedings of the IEEE/CVF Conference on Computer Vision
  and Pattern Recognition}, pages 10136--10145, 2020{\natexlab{b}}.

\bibitem[Bettaney et~al.(2019)Bettaney, Hardwick, Zisimopoulos, and
  Chamberlain]{bettaney2019fashion}
Elaine~M Bettaney, Stephen~R Hardwick, Odysseas Zisimopoulos, and Benjamin~Paul
  Chamberlain.
\newblock Fashion outfit generation for e-commerce.
\newblock \emph{arXiv preprint arXiv:1904.00741}, 2019.

\bibitem[Bossard et~al.(2012)Bossard, Dantone, Leistner, Wengert, Quack, and
  Van~Gool]{bossard2012apparel}
Lukas Bossard, Matthias Dantone, Christian Leistner, Christian Wengert, Till
  Quack, and Luc Van~Gool.
\newblock Apparel classification with style.
\newblock In \emph{Asian conference on computer vision}, pages 321--335.
  Springer, 2012.

\bibitem[Caliskan et~al.(2020)Caliskan, Mustafa, Imre, and
  Hilton]{caliskan2020multi}
Akin Caliskan, Armin Mustafa, Evren Imre, and Adrian Hilton.
\newblock Multi-view consistency loss for improved single-image 3d
  reconstruction of clothed people.
\newblock In \emph{Proceedings of the Asian Conference on Computer Vision},
  2020.

\bibitem[Castro and Ramirez(2020)]{castro2020segmentation}
Hassler Castro and Mariana Ramirez.
\newblock Segmentation task for fashion and apparel.
\newblock \emph{arXiv preprint arXiv:2006.11375}, 2020.

\bibitem[Che et~al.(2016)Che, Li, Jacob, Bengio, and Li]{che2016mode}
Tong Che, Yanran Li, Athul~Paul Jacob, Yoshua Bengio, and Wenjie Li.
\newblock Mode regularized generative adversarial networks.
\newblock \emph{arXiv preprint arXiv:1612.02136}, 2016.

\bibitem[Chen et~al.(2012)Chen, Gallagher, and Girod]{chen2012describing}
Huizhong Chen, Andrew Gallagher, and Bernd Girod.
\newblock Describing clothing by semantic attributes.
\newblock In \emph{European conference on computer vision}, pages 609--623.
  Springer, 2012.

\bibitem[Chen et~al.(2020)Chen, Tian, Li, Wu, King, Chen, Hsieh, and
  Xu]{chen2020tailorgan}
Lele Chen, Justin Tian, Guo Li, Cheng-Haw Wu, Erh-Kan King, Kuan-Ting Chen,
  Shao-Hang Hsieh, and Chenliang Xu.
\newblock Tailorgan: Making user-defined fashion designs.
\newblock In \emph{Proceedings of the IEEE/CVF Winter Conference on
  Applications of Computer Vision}, pages 3241--3250, 2020.

\bibitem[Chen et~al.(2019)Chen, Huang, Xu, Guo, Guo, Sun, Li, Pfadler, Zhao,
  and Zhao]{chen2019pog}
Wen Chen, Pipei Huang, Jiaming Xu, Xin Guo, Cheng Guo, Fei Sun, Chao Li,
  Andreas Pfadler, Huan Zhao, and Binqiang Zhao.
\newblock Pog: personalized outfit generation for fashion recommendation at
  alibaba ifashion.
\newblock In \emph{Proceedings of the 25th ACM SIGKDD international conference
  on knowledge discovery \& data mining}, pages 2662--2670, 2019.

\bibitem[Cheng et~al.(2020)Cheng, Song, Chen, Hidayati, and
  Liu]{cheng2020fashion}
Wen-Huang Cheng, Sijie Song, Chieh-Yun Chen, Shintami~Chusnul Hidayati, and
  Jiaying Liu.
\newblock Fashion meets computer vision: A survey.
\newblock \emph{arXiv preprint arXiv:2003.13988}, 2020.

\bibitem[Cheng et~al.(2017)Cheng, Wu, Liu, and Hua]{cheng2017video2shop}
Zhi-Qi Cheng, Xiao Wu, Yang Liu, and Xian-Sheng Hua.
\newblock Video2shop: Exact matching clothes in videos to online shopping
  images.
\newblock In \emph{Proceedings of the IEEE Conference on Computer Vision and
  Pattern Recognition}, pages 4048--4056, 2017.

\bibitem[Chou et~al.(2018)Chou, Lee, Zhang, Lee, and Hsu]{chou2018pivtons}
Chao-Te Chou, Cheng-Han Lee, Kaipeng Zhang, Hu-Cheng Lee, and Winston~H Hsu.
\newblock Pivtons: Pose invariant virtual try-on shoe with conditional image
  completion.
\newblock In \emph{Asian Conference on Computer Vision}, pages 654--668.
  Springer, 2018.

\bibitem[Cucurull et~al.(2019)Cucurull, Taslakian, and
  Vazquez]{cucurull2019context}
Guillem Cucurull, Perouz Taslakian, and David Vazquez.
\newblock Context-aware visual compatibility prediction.
\newblock In \emph{Proceedings of the IEEE/CVF Conference on Computer Vision
  and Pattern Recognition}, pages 12617--12626, 2019.

\bibitem[Cui et~al.(2018)Cui, Liu, Gao, and Su]{cui2018fashiongan}
Yi~Rui Cui, Qi~Liu, Cheng~Ying Gao, and Zhongbo Su.
\newblock Fashiongan: Display your fashion design using conditional generative
  adversarial nets.
\newblock In \emph{Computer Graphics Forum}, volume~37, pages 109--119. Wiley
  Online Library, 2018.

\bibitem[Cui et~al.(2019)Cui, Li, Wu, Zhang, and Wang]{cui2019dressing}
Zeyu Cui, Zekun Li, Shu Wu, Xiao-Yu Zhang, and Liang Wang.
\newblock Dressing as a whole: Outfit compatibility learning based on node-wise
  graph neural networks.
\newblock In \emph{The World Wide Web Conference}, pages 307--317, 2019.

\bibitem[Date et~al.(2017)Date, Ganesan, and Oates]{date2017fashioning}
Prutha Date, Ashwinkumar Ganesan, and Tim Oates.
\newblock Fashioning with networks: neural style transfer to design clothes.
\newblock In \emph{KDD ML4Fashion workshop}, 2017.

\bibitem[Di et~al.(2013)Di, Wah, Bhardwaj, Piramuthu, and
  Sundaresan]{di2013style}
Wei Di, Catherine Wah, Anurag Bhardwaj, Robinson Piramuthu, and Neel
  Sundaresan.
\newblock Style finder: Fine-grained clothing style detection and retrieval.
\newblock In \emph{Proceedings of the IEEE Conference on computer vision and
  pattern recognition workshops}, pages 8--13, 2013.

\bibitem[Dong et~al.(2018{\natexlab{a}})Dong, Liang, Gong, Lai, Zhu, and
  Yin]{dong2018soft}
Haoye Dong, Xiaodan Liang, Ke~Gong, Hanjiang Lai, Jia Zhu, and Jian Yin.
\newblock Soft-gated warping-gan for pose-guided person image synthesis.
\newblock \emph{arXiv preprint arXiv:1810.11610}, 2018{\natexlab{a}}.

\bibitem[Dong et~al.(2019{\natexlab{a}})Dong, Liang, Shen, Wang, Lai, Zhu, Hu,
  and Yin]{dong2019towards}
Haoye Dong, Xiaodan Liang, Xiaohui Shen, Bochao Wang, Hanjiang Lai, Jia Zhu,
  Zhiting Hu, and Jian Yin.
\newblock Towards multi-pose guided virtual try-on network.
\newblock In \emph{Proceedings of the IEEE/CVF International Conference on
  Computer Vision}, pages 9026--9035, 2019{\natexlab{a}}.

\bibitem[Dong et~al.(2019{\natexlab{b}})Dong, Liang, Shen, Wu, Chen, and
  Yin]{dong2019fw}
Haoye Dong, Xiaodan Liang, Xiaohui Shen, Bowen Wu, Bing-Cheng Chen, and Jian
  Yin.
\newblock Fw-gan: Flow-navigated warping gan for video virtual try-on.
\newblock In \emph{Proceedings of the IEEE/CVF International Conference on
  Computer Vision}, pages 1161--1170, 2019{\natexlab{b}}.

\bibitem[Dong et~al.(2019{\natexlab{c}})Dong, Liang, Zhang, Zhang, Xie, Wu,
  Zhang, Shen, and Yin]{dong2019fashion}
Haoye Dong, Xiaodan Liang, Yixuan Zhang, Xujie Zhang, Zhenyu Xie, Bowen Wu,
  Ziqi Zhang, Xiaohui Shen, and Jian Yin.
\newblock Fashion editing with multi-scale attention normalization.
\newblock \emph{CoRR, abs/1906.00884}, 2019{\natexlab{c}}.

\bibitem[Dong et~al.(2020)Dong, Liang, Zhang, Zhang, Shen, Xie, Wu, and
  Yin]{dong2020fashion}
Haoye Dong, Xiaodan Liang, Yixuan Zhang, Xujie Zhang, Xiaohui Shen, Zhenyu Xie,
  Bowen Wu, and Jian Yin.
\newblock Fashion editing with adversarial parsing learning.
\newblock In \emph{Proceedings of the IEEE/CVF Conference on Computer Vision
  and Pattern Recognition}, pages 8120--8128, 2020.

\bibitem[Dong et~al.(2013)Dong, Chen, Xia, Huang, and Yan]{dong2013deformable}
Jian Dong, Qiang Chen, Wei Xia, Zhongyang Huang, and Shuicheng Yan.
\newblock A deformable mixture parsing model with parselets.
\newblock In \emph{Proceedings of the IEEE International Conference on Computer
  Vision}, pages 3408--3415, 2013.

\bibitem[Dong et~al.(2014)Dong, Chen, Shen, Yang, and Yan]{dong2014towards}
Jian Dong, Qiang Chen, Xiaohui Shen, Jianchao Yang, and Shuicheng Yan.
\newblock Towards unified human parsing and pose estimation.
\newblock In \emph{Proceedings of the IEEE Conference on Computer Vision and
  Pattern Recognition}, pages 843--850, 2014.

\bibitem[Dong et~al.(2015)Dong, Chen, Huang, Yang, and Yan]{dong2015parsing}
Jian Dong, Qiang Chen, Zhongyang Huang, Jianchao Yang, and Shuicheng Yan.
\newblock Parsing based on parselets: A unified deformable mixture model for
  human parsing.
\newblock \emph{IEEE transactions on pattern analysis and machine
  intelligence}, 38\penalty0 (1):\penalty0 88--101, 2015.

\bibitem[Dong et~al.(2017{\natexlab{a}})Dong, Gong, and Zhu]{dong2017class}
Qi~Dong, Shaogang Gong, and Xiatian Zhu.
\newblock Class rectification hard mining for imbalanced deep learning.
\newblock In \emph{Proceedings of the IEEE International Conference on Computer
  Vision}, pages 1851--1860, 2017{\natexlab{a}}.

\bibitem[Dong et~al.(2017{\natexlab{b}})Dong, Gong, and Zhu]{dong2017multi}
Qi~Dong, Shaogang Gong, and Xiatian Zhu.
\newblock Multi-task curriculum transfer deep learning of clothing attributes.
\newblock In \emph{2017 IEEE Winter Conference on Applications of Computer
  Vision (WACV)}, pages 520--529. IEEE, 2017{\natexlab{b}}.

\bibitem[Dong et~al.(2018{\natexlab{b}})Dong, Gong, and
  Zhu]{dong2018imbalanced}
Qi~Dong, Shaogang Gong, and Xiatian Zhu.
\newblock Imbalanced deep learning by minority class incremental rectification.
\newblock \emph{IEEE transactions on pattern analysis and machine
  intelligence}, 41\penalty0 (6):\penalty0 1367--1381, 2018{\natexlab{b}}.

\bibitem[Dubey et~al.(2020)Dubey, Bhardwaj, Abhinav, Kuriakose, Jain, and
  Arora]{dubey2020ai}
Alpana Dubey, Nitish Bhardwaj, Kumar Abhinav, Suma~Mani Kuriakose, Sakshi Jain,
  and Veenu Arora.
\newblock Ai assisted apparel design.
\newblock \emph{arXiv preprint arXiv:2007.04950}, 2020.

\bibitem[Esser et~al.(2018)Esser, Sutter, and Ommer]{esser2018variational}
Patrick Esser, Ekaterina Sutter, and Bj{\"o}rn Ommer.
\newblock A variational u-net for conditional appearance and shape generation.
\newblock In \emph{Proceedings of the IEEE Conference on Computer Vision and
  Pattern Recognition}, pages 8857--8866, 2018.

\bibitem[Ge et~al.(2019)Ge, Zhang, Wang, Tang, and Luo]{ge2019deepfashion2}
Yuying Ge, Ruimao Zhang, Xiaogang Wang, Xiaoou Tang, and Ping Luo.
\newblock Deepfashion2: A versatile benchmark for detection, pose estimation,
  segmentation and re-identification of clothing images.
\newblock In \emph{Proceedings of the IEEE/CVF Conference on Computer Vision
  and Pattern Recognition}, pages 5337--5345, 2019.

\bibitem[Gong and Khalid(2021)]{gong2021aesthetics}
Wei Gong and Laila Khalid.
\newblock Aesthetics, personalization and recommendation: A survey on deep
  learning in fashion.
\newblock \emph{arXiv preprint arXiv:2101.08301}, 2021.

\bibitem[Gu et~al.(2019)Gu, Wang, Chiu, Tai, and Tang]{gu2019ladn}
Qiao Gu, Guanzhi Wang, Mang~Tik Chiu, Yu-Wing Tai, and Chi-Keung Tang.
\newblock Ladn: Local adversarial disentangling network for facial makeup and
  de-makeup.
\newblock In \emph{Proceedings of the IEEE/CVF International Conference on
  Computer Vision}, pages 10481--10490, 2019.

\bibitem[Gu et~al.(2017)Gu, Wong, Peng, Shou, Chen, and
  Kankanhalli]{gu2017understanding}
Xiaoling Gu, Yongkang Wong, Pai Peng, Lidan Shou, Gang Chen, and Mohan~S
  Kankanhalli.
\newblock Understanding fashion trends from street photos via
  neighbor-constrained embedding learning.
\newblock In \emph{Proceedings of the 25th ACM international conference on
  Multimedia}, pages 190--198, 2017.

\bibitem[Gu et~al.(2018)Gu, Wong, Shou, Peng, Chen, and
  Kankanhalli]{gu2018multi}
Xiaoling Gu, Yongkang Wong, Lidan Shou, Pai Peng, Gang Chen, and Mohan~S
  Kankanhalli.
\newblock Multi-modal and multi-domain embedding learning for fashion retrieval
  and analysis.
\newblock \emph{IEEE Transactions on Multimedia}, 21\penalty0 (6):\penalty0
  1524--1537, 2018.

\bibitem[Gundogdu et~al.(2019)Gundogdu, Constantin, Seifoddini, Dang, Salzmann,
  and Fua]{gundogdu2019garnet}
Erhan Gundogdu, Victor Constantin, Amrollah Seifoddini, Minh Dang, Mathieu
  Salzmann, and Pascal Fua.
\newblock Garnet: A two-stream network for fast and accurate 3d cloth draping.
\newblock In \emph{Proceedings of the IEEE/CVF International Conference on
  Computer Vision}, pages 8739--8748, 2019.

\bibitem[Gundogdu et~al.(2020)Gundogdu, Constantin, Parashar, Banadkooki, Dang,
  Salzmann, and Fua]{gundogdu2020garnet++}
Erhan Gundogdu, Victor Constantin, Shaifali Parashar, Amrollah~Seifoddini
  Banadkooki, Minh Dang, Mathieu Salzmann, and Pascal Fua.
\newblock Garnet++: Improving fast and accurate static 3d cloth draping by
  curvature loss.
\newblock \emph{IEEE Transactions on Pattern Analysis and Machine
  Intelligence}, 2020.

\bibitem[Guo et~al.(2019)Guo, Huang, Zhang, Srikhanta, Cui, Li, Adam, Scott,
  and Belongie]{guo2019imaterialist}
Sheng Guo, Weilin Huang, Xiao Zhang, Prasanna Srikhanta, Yin Cui, Yuan Li,
  Hartwig Adam, Matthew~R Scott, and Serge Belongie.
\newblock The imaterialist fashion attribute dataset.
\newblock In \emph{Proceedings of the IEEE International Conference on Computer
  Vision Workshops}, pages 0--0, 2019.

\bibitem[Hadi~Kiapour and Piramuthu(2018)]{hadi2018brand}
M~Hadi~Kiapour and Robinson Piramuthu.
\newblock Brand> logo: Visual analysis of fashion brands.
\newblock In \emph{Proceedings of the European Conference on Computer Vision
  (ECCV)}, pages 0--0, 2018.

\bibitem[Hadi~Kiapour et~al.(2015)Hadi~Kiapour, Han, Lazebnik, Berg, and
  Berg]{hadi2015buy}
M~Hadi~Kiapour, Xufeng Han, Svetlana Lazebnik, Alexander~C Berg, and Tamara~L
  Berg.
\newblock Where to buy it: Matching street clothing photos in online shops.
\newblock In \emph{Proceedings of the IEEE international conference on computer
  vision}, pages 3343--3351, 2015.

\bibitem[Han and Zhu(2008)]{han2008bottom}
Feng Han and Song-Chun Zhu.
\newblock Bottom-up/top-down image parsing with attribute grammar.
\newblock \emph{IEEE Transactions on Pattern Analysis and Machine
  Intelligence}, 31\penalty0 (1):\penalty0 59--73, 2008.

\bibitem[Han et~al.(2017{\natexlab{a}})Han, Wu, Huang, Zhang, Zhu, Li, Zhao,
  and Davis]{han2017automatic}
Xintong Han, Zuxuan Wu, Phoenix~X. Huang, Xiao Zhang, Menglong Zhu, Yuan Li,
  Yang Zhao, and Larry~S. Davis.
\newblock Automatic spatially-aware fashion concept discovery.
\newblock In \emph{ICCV}, 2017{\natexlab{a}}.

\bibitem[Han et~al.(2017{\natexlab{b}})Han, Wu, Jiang, and
  Davis]{han2017learning}
Xintong Han, Zuxuan Wu, Yu-Gang Jiang, and Larry~S Davis.
\newblock Learning fashion compatibility with bidirectional lstms.
\newblock In \emph{Proceedings of the 25th ACM international conference on
  Multimedia}, pages 1078--1086, 2017{\natexlab{b}}.

\bibitem[Han et~al.(2018)Han, Wu, Wu, Yu, and Davis]{han2018viton}
Xintong Han, Zuxuan Wu, Zhe Wu, Ruichi Yu, and Larry~S Davis.
\newblock Viton: An image-based virtual try-on network.
\newblock In \emph{Proceedings of the IEEE conference on computer vision and
  pattern recognition}, pages 7543--7552, 2018.

\bibitem[Han et~al.(2019{\natexlab{a}})Han, Hu, Huang, and
  Scott]{han2019clothflow}
Xintong Han, Xiaojun Hu, Weilin Huang, and Matthew~R Scott.
\newblock Clothflow: A flow-based model for clothed person generation.
\newblock In \emph{Proceedings of the IEEE/CVF International Conference on
  Computer Vision}, pages 10471--10480, 2019{\natexlab{a}}.

\bibitem[Han et~al.(2019{\natexlab{b}})Han, Wu, Huang, Scott, and
  Davis]{han2019finet}
Xintong Han, Zuxuan Wu, Weilin Huang, Matthew~R Scott, and Larry~S Davis.
\newblock Finet: Compatible and diverse fashion image inpainting.
\newblock In \emph{Proceedings of the IEEE/CVF International Conference on
  Computer Vision}, pages 4481--4491, 2019{\natexlab{b}}.

\bibitem[Han et~al.(2020)Han, Yang, Wang, and Liu]{han2020design}
Yu~Han, Shuai Yang, Wenjing Wang, and Jiaying Liu.
\newblock From design draft to real attire: Unaligned fashion image
  translation.
\newblock In \emph{Proceedings of the 28th ACM International Conference on
  Multimedia}, pages 1533--1541, 2020.

\bibitem[Hazra~Banerjee et~al.(2020)Hazra~Banerjee, Ravi, and
  Dutta]{hazra2020attr2style}
Rajdeep Hazra~Banerjee, Abhinav Ravi, and Ujjal~Kr Dutta.
\newblock Attr2style: A transfer learning approach for inferring fashion styles
  via apparel attributes.
\newblock \emph{arXiv e-prints}, pages arXiv--2008, 2020.

\bibitem[He and McAuley(2016)]{he2016ups}
Ruining He and Julian McAuley.
\newblock Ups and downs: Modeling the visual evolution of fashion trends with
  one-class collaborative filtering.
\newblock In \emph{proceedings of the 25th international conference on world
  wide web}, pages 507--517, 2016.

\bibitem[He et~al.(2016)He, Packer, and McAuley]{he2016learning}
Ruining He, Charles Packer, and Julian McAuley.
\newblock Learning compatibility across categories for heterogeneous item
  recommendation.
\newblock In \emph{2016 IEEE 16th International Conference on Data Mining
  (ICDM)}, pages 937--942. IEEE, 2016.

\bibitem[Hidayati et~al.(2020)Hidayati, Goh, Chan, Hsu, See, Kuan, Hua, Tsao,
  and Cheng]{hidayati2020dress}
Shintami~Chusnul Hidayati, Ting~Wei Goh, Ji-Sheng~Gary Chan, Cheng-Chun Hsu,
  John See, Wong~Lai Kuan, Kai-Lung Hua, Yu~Tsao, and Wen-Huang Cheng.
\newblock Dress with style: Learning style from joint deep embedding of
  clothing styles and body shapes.
\newblock \emph{IEEE Transactions on Multimedia}, 2020.

\bibitem[Hsiao and Grauman(2021)]{hsiao2021culture}
Wei-Lin Hsiao and Kristen Grauman.
\newblock From culture to clothing: Discovering the world events behind a
  century of fashion images.
\newblock \emph{arXiv preprint arXiv:2102.01690}, 2021.

\bibitem[Hsiao et~al.(2019)Hsiao, Katsman, Wu, Parikh, and
  Grauman]{hsiao2019fashion++}
Wei-Lin Hsiao, Isay Katsman, Chao-Yuan Wu, Devi Parikh, and Kristen Grauman.
\newblock Fashion++: Minimal edits for outfit improvement.
\newblock In \emph{Proceedings of the IEEE/CVF International Conference on
  Computer Vision}, pages 5047--5056, 2019.

\bibitem[Hu et~al.(2018)Hu, Yang, Salakhutdinov, Liang, Qin, Dong, and
  Xing]{hu2018deep}
Zhiting Hu, Zichao Yang, Ruslan Salakhutdinov, Xiaodan Liang, Lianhui Qin,
  Haoye Dong, and Eric Xing.
\newblock Deep generative models with learnable knowledge constraints.
\newblock \emph{arXiv preprint arXiv:1806.09764}, 2018.

\bibitem[Huang et~al.(2015)Huang, Feris, Chen, and Yan]{huang2015cross}
Junshi Huang, Rogerio~S Feris, Qiang Chen, and Shuicheng Yan.
\newblock Cross-domain image retrieval with a dual attribute-aware ranking
  network.
\newblock In \emph{Proceedings of the IEEE international conference on computer
  vision}, pages 1062--1070, 2015.

\bibitem[Huang et~al.(2020)Huang, Xu, Lassner, Li, and Tung]{huang2020arch}
Zeng Huang, Yuanlu Xu, Christoph Lassner, Hao Li, and Tony Tung.
\newblock Arch: Animatable reconstruction of clothed humans.
\newblock In \emph{Proceedings of the IEEE/CVF Conference on Computer Vision
  and Pattern Recognition}, pages 3093--3102, 2020.

\bibitem[Inoue et~al.(2017)Inoue, Simo-Serra, Yamasaki, and
  Ishikawa]{InoueICCVW2017}
Naoto Inoue, Edgar Simo-Serra, Toshihiko Yamasaki, and Hiroshi Ishikawa.
\newblock {Multi-Label Fashion Image Classification with Minimal Human
  Supervision}.
\newblock In \emph{Proceedings of the International Conference on Computer
  Vision Workshops (ICCVW)}, 2017.

\bibitem[Jagadeesh et~al.(2014)Jagadeesh, Piramuthu, Bhardwaj, Di, and
  Sundaresan]{jagadeesh2014large}
Vignesh Jagadeesh, Robinson Piramuthu, Anurag Bhardwaj, Wei Di, and Neel
  Sundaresan.
\newblock Large scale visual recommendations from street fashion images.
\newblock In \emph{Proceedings of the 20th ACM SIGKDD international conference
  on Knowledge discovery and data mining}, pages 1925--1934, 2014.

\bibitem[Jia et~al.(2019)Jia, Shi, Sirotenko, Cui, Hariharan, Cardie, and
  Belongie]{jia2019fashionpedia}
Menglin Jia, Mengyun Shi, Mikhail Sirotenko, Yin Cui, Bharath Hariharan, Claire
  Cardie, and Serge Belongie.
\newblock The fashionpedia ontology and fashion segmentation dataset.
\newblock \emph{Cornell University}, 2019.

\bibitem[Jiang et~al.(2020)Jiang, Zhang, Hong, Luo, Liu, and
  Bao]{jiang2020bcnet}
Boyi Jiang, Juyong Zhang, Yang Hong, Jinhao Luo, Ligang Liu, and Hujun Bao.
\newblock Bcnet: Learning body and cloth shape from a single image.
\newblock In \emph{European Conference on Computer Vision}, pages 18--35.
  Springer, 2020.

\bibitem[Jiang and Fu(2017)]{jiang2017fashion}
Shuhui Jiang and Yun Fu.
\newblock Fashion style generator.
\newblock In \emph{Proceedings of the 26th International Joint Conference on
  Artificial Intelligence}, pages 3721--3727, 2017.

\bibitem[Kang et~al.(2017)Kang, Fang, Wang, and McAuley]{kang2017visually}
Wang-Cheng Kang, Chen Fang, Zhaowen Wang, and Julian McAuley.
\newblock Visually-aware fashion recommendation and design with generative
  image models.
\newblock In \emph{2017 IEEE International Conference on Data Mining (ICDM)},
  pages 207--216. IEEE, 2017.

\bibitem[Kang et~al.(2019)Kang, Kim, Leskovec, Rosenberg, and
  McAuley]{kang2019complete}
Wang-Cheng Kang, Eric Kim, Jure Leskovec, Charles Rosenberg, and Julian
  McAuley.
\newblock Complete the look: Scene-based complementary product recommendation.
\newblock In \emph{Proceedings of the IEEE/CVF Conference on Computer Vision
  and Pattern Recognition}, pages 10532--10541, 2019.

\bibitem[Kashilani et~al.(2018)Kashilani, Damahe, and
  Thakur]{kashilani2018overview}
Divva Kashilani, Lalit~B Damahe, and Nileshsingh~V Thakur.
\newblock An overview of image recognition and retrieval of clothing items.
\newblock In \emph{2018 International Conference on Research in Intelligent and
  Computing in Engineering (RICE)}, pages 1--6. IEEE, 2018.

\bibitem[Kato et~al.(2018)Kato, Osone, Sato, Muramatsu, and
  Ochiai]{kato2018deepwear}
Natsumi Kato, Hiroyuki Osone, Daitetsu Sato, Naoya Muramatsu, and Yoichi
  Ochiai.
\newblock Deepwear: a case study of collaborative design between human and
  artificial intelligence.
\newblock In \emph{Proceedings of the Twelfth International Conference on
  Tangible, Embedded, and Embodied Interaction}, pages 529--536, 2018.

\bibitem[Kato et~al.(2019)Kato, Osone, Oomori, Ooi, and Ochiai]{kato2019gans}
Natsumi Kato, Hiroyuki Osone, Kotaro Oomori, Chun~Wei Ooi, and Yoichi Ochiai.
\newblock Gans-based clothes design: Pattern maker is all you need to design
  clothing.
\newblock In \emph{Proceedings of the 10th Augmented Human International
  Conference 2019}, pages 1--7, 2019.

\bibitem[Kiapour et~al.(2014)Kiapour, Yamaguchi, Berg, and
  Berg]{kiapour2014hipster}
M~Hadi Kiapour, Kota Yamaguchi, Alexander~C Berg, and Tamara~L Berg.
\newblock Hipster wars: Discovering elements of fashion styles.
\newblock In \emph{European conference on computer vision}, pages 472--488.
  Springer, 2014.

\bibitem[Kim et~al.(2020)Kim, Saito, Saenko, Sclaroff, and
  Plummer]{kim2020self}
Donghyun Kim, Kuniaki Saito, Kate Saenko, Stan Sclaroff, and Bryan~A Plummer.
\newblock Self-supervised visual attribute learning for fashion compatibility.
\newblock \emph{arXiv preprint arXiv:2008.00348}, 2020.

\bibitem[Kovashka et~al.(2012)Kovashka, Parikh, and
  Grauman]{kovashka2012whittlesearch}
Adriana Kovashka, Devi Parikh, and Kristen Grauman.
\newblock Whittlesearch: Image search with relative attribute feedback.
\newblock In \emph{2012 IEEE Conference on Computer Vision and Pattern
  Recognition}, pages 2973--2980. IEEE, 2012.

\bibitem[Kuang et~al.(2018)Kuang, Yu, Yu, and Fan]{kuang2018ontology}
Zhenzhong Kuang, Jun Yu, Zhou Yu, and Jianping Fan.
\newblock Ontology-driven hierarchical deep learning for fashion recognition.
\newblock In \emph{2018 IEEE Conference on Multimedia Information Processing
  and Retrieval (MIPR)}, pages 19--24. IEEE, 2018.

\bibitem[Kuppa et~al.(2020)Kuppa, Jong, Liu, Liu, and Moh]{kuppa2020shineon}
Gaurav Kuppa, Andrew Jong, Xin Liu, Ziwei Liu, and Teng-Sheng Moh.
\newblock Shineon: Illuminating design choices for practical video-based
  virtual clothing try-on.
\newblock In \emph{Proceedings of the IEEE/CVF Winter Conference on
  Applications of Computer Vision}, pages 191--200, 2020.

\bibitem[Laenen and Moens(2020)]{laenen2020attention}
Katrien Laenen and Marie-Francine Moens.
\newblock Attention-based fusion for outfit recommendation.
\newblock In \emph{Fashion Recommender Systems}, pages 69--86. Springer, 2020.

\bibitem[Laenen et~al.(2017)Laenen, Zoghbi, and Moens]{laenen2017cross}
Katrien Laenen, Susana Zoghbi, and Marie-Francine Moens.
\newblock Cross-modal search for fashion attributes.
\newblock In \emph{Proceedings of the KDD 2017 Workshop on Machine Learning
  Meets Fashion}, volume 2017, pages 1--10. ACM, 2017.

\bibitem[Lee et~al.(2019)Lee, Oh, Jung, and Kim]{lee2019global}
Sumin Lee, Sungchan Oh, Chanho Jung, and Changick Kim.
\newblock A global-local embedding module for fashion landmark detection.
\newblock In \emph{Proceedings of the IEEE/CVF International Conference on
  Computer Vision Workshops}, pages 0--0, 2019.

\bibitem[Li et~al.(2019{\natexlab{a}})Li, Liu, and Forsyth]{li2019coherent}
Kedan Li, Chen Liu, and David Forsyth.
\newblock Coherent and controllable outfit generation.
\newblock \emph{arXiv preprint arXiv:1906.07273}, 2019{\natexlab{a}}.

\bibitem[Li et~al.(2017)Li, Arnab, and Torr]{li2017holistic}
Qizhu Li, Anurag Arnab, and Philip~HS Torr.
\newblock Holistic, instance-level human parsing.
\newblock \emph{arXiv preprint arXiv:1709.03612}, 2017.

\bibitem[Li et~al.(2018)Li, Qian, Dong, Liu, Yan, Zhu, and
  Lin]{li2018beautygan}
Tingting Li, Ruihe Qian, Chao Dong, Si~Liu, Qiong Yan, Wenwu Zhu, and Liang
  Lin.
\newblock Beautygan: Instance-level facial makeup transfer with deep generative
  adversarial network.
\newblock In \emph{Proceedings of the 26th ACM international conference on
  Multimedia}, pages 645--653, 2018.

\bibitem[Li et~al.(2020{\natexlab{a}})Li, Wang, He, Chen, Xiao, and
  Chua]{li2020hierarchical}
Xingchen Li, Xiang Wang, Xiangnan He, Long Chen, Jun Xiao, and Tat-Seng Chua.
\newblock Hierarchical fashion graph network for personalized outfit
  recommendation.
\newblock In \emph{Proceedings of the 43rd International ACM SIGIR Conference
  on Research and Development in Information Retrieval}, pages 159--168,
  2020{\natexlab{a}}.

\bibitem[Li et~al.(2020{\natexlab{b}})Li, Yu, Han, Jiang, Jia, and
  Lu]{li2020deep1}
Yao Li, Xianggang Yu, Xiaoguang Han, Nianjuan Jiang, Kui Jia, and Jiangbo Lu.
\newblock A deep learning based interactive sketching system for fashion images
  design.
\newblock \emph{arXiv preprint arXiv:2010.04413}, 2020{\natexlab{b}}.

\bibitem[Li et~al.(2019{\natexlab{b}})Li, Tang, Ye, and Ma]{li2019spatial}
Yixin Li, Shengqin Tang, Yun Ye, and Jinwen Ma.
\newblock Spatial-aware non-local attention for fashion landmark detection.
\newblock In \emph{2019 IEEE International Conference on Multimedia and Expo
  (ICME)}, pages 820--825. IEEE, 2019{\natexlab{b}}.

\bibitem[Li et~al.(2020{\natexlab{c}})Li, Habermann, Thomaszewski, Coros,
  Beeler, and Theobalt]{li2020deep}
Yue Li, Marc Habermann, Bernhard Thomaszewski, Stelian Coros, Thabo Beeler, and
  Christian Theobalt.
\newblock Deep physics-aware inference of cloth deformation for monocular human
  performance capture.
\newblock \emph{arXiv preprint arXiv:2011.12866}, 2020{\natexlab{c}}.

\bibitem[Liang et~al.(2015{\natexlab{a}})Liang, Liu, Shen, Yang, Liu, Dong,
  Lin, and Yan]{liang2015deep}
Xiaodan Liang, Si~Liu, Xiaohui Shen, Jianchao Yang, Luoqi Liu, Jian Dong, Liang
  Lin, and Shuicheng Yan.
\newblock Deep human parsing with active template regression.
\newblock \emph{IEEE transactions on pattern analysis and machine
  intelligence}, 37\penalty0 (12):\penalty0 2402--2414, 2015{\natexlab{a}}.

\bibitem[Liang et~al.(2015{\natexlab{b}})Liang, Xu, Shen, Yang, Liu, Tang, Lin,
  and Yan]{liang2015human}
Xiaodan Liang, Chunyan Xu, Xiaohui Shen, Jianchao Yang, Si~Liu, Jinhui Tang,
  Liang Lin, and Shuicheng Yan.
\newblock Human parsing with contextualized convolutional neural network.
\newblock In \emph{Proceedings of the IEEE international conference on computer
  vision}, pages 1386--1394, 2015{\natexlab{b}}.

\bibitem[Liang et~al.(2016{\natexlab{a}})Liang, Lin, Yang, Luo, Huang, and
  Yan]{liang2016clothes}
Xiaodan Liang, Liang Lin, Wei Yang, Ping Luo, Junshi Huang, and Shuicheng Yan.
\newblock Clothes co-parsing via joint image segmentation and labeling with
  application to clothing retrieval.
\newblock \emph{IEEE Transactions on Multimedia}, 18\penalty0 (6):\penalty0
  1175--1186, 2016{\natexlab{a}}.

\bibitem[Liang et~al.(2016{\natexlab{b}})Liang, Shen, Feng, Lin, and
  Yan]{liang2016semantic}
Xiaodan Liang, Xiaohui Shen, Jiashi Feng, Liang Lin, and Shuicheng Yan.
\newblock Semantic object parsing with graph lstm.
\newblock In \emph{European Conference on Computer Vision}, pages 125--143.
  Springer, 2016{\natexlab{b}}.

\bibitem[Liang et~al.(2018)Liang, Gong, Shen, and Lin]{liang2018look}
Xiaodan Liang, Ke~Gong, Xiaohui Shen, and Liang Lin.
\newblock Look into person: Joint body parsing \& pose estimation network and a
  new benchmark.
\newblock \emph{IEEE transactions on pattern analysis and machine
  intelligence}, 41\penalty0 (4):\penalty0 871--885, 2018.

\bibitem[Lin et~al.(2015)Lin, Yang, Liu, Hsiao, and Chen]{lin2015rapid}
Kevin Lin, Huei-Fang Yang, Kuan-Hsien Liu, Jen-Hao Hsiao, and Chu-Song Chen.
\newblock Rapid clothing retrieval via deep learning of binary codes and
  hierarchical search.
\newblock In \emph{Proceedings of the 5th ACM on International Conference on
  Multimedia Retrieval}, pages 499--502, 2015.

\bibitem[Lin(2020)]{lin2020aggregation}
Tzu-Heng Lin.
\newblock Aggregation and finetuning for clothes landmark detection.
\newblock \emph{arXiv preprint arXiv:2005.00419}, 2020.

\bibitem[Lin et~al.(2020)Lin, Tran, and Davis]{lin2020fashion}
Yen-Liang Lin, Son Tran, and Larry~S Davis.
\newblock Fashion outfit complementary item retrieval.
\newblock In \emph{Proceedings of the IEEE/CVF Conference on Computer Vision
  and Pattern Recognition}, pages 3311--3319, 2020.

\bibitem[Lin et~al.(2019)Lin, Moosaei, and Yang]{lin2019learning}
Yusan Lin, Maryam Moosaei, and Hao Yang.
\newblock Learning personal tastes in choosing fashion outfits.
\newblock In \emph{Proceedings of the IEEE/CVF Conference on Computer Vision
  and Pattern Recognition Workshops}, pages 0--0, 2019.

\bibitem[Liu et~al.(2016{\natexlab{a}})Liu, Chen, and Chen]{liu2016mvc}
Kuan-Hsien Liu, Ting-Yen Chen, and Chu-Song Chen.
\newblock Mvc: A dataset for view-invariant clothing retrieval and attribute
  prediction.
\newblock In \emph{Proceedings of the 2016 ACM on International Conference on
  Multimedia Retrieval}, pages 313--316, 2016{\natexlab{a}}.

\bibitem[Liu et~al.(2019{\natexlab{a}})Liu, Zhang, Ji, and Wu]{liu2019toward}
Linlin Liu, Haijun Zhang, Yuzhu Ji, and QM~Jonathan Wu.
\newblock Toward ai fashion design: An attribute-gan model for clothing match.
\newblock \emph{Neurocomputing}, 341:\penalty0 156--167, 2019{\natexlab{a}}.

\bibitem[Liu et~al.(2011)Liu, Yan, Zhang, Xu, Liu, and Lu]{liu2011weakly}
Si~Liu, Shuicheng Yan, Tianzhu Zhang, Changsheng Xu, Jing Liu, and Hanqing Lu.
\newblock Weakly supervised graph propagation towards collective image parsing.
\newblock \emph{IEEE Transactions on Multimedia}, 14\penalty0 (2):\penalty0
  361--373, 2011.

\bibitem[Liu et~al.(2012{\natexlab{a}})Liu, Feng, Song, Zhang, Lu, Xu, and
  Yan]{liu2012hi}
Si~Liu, Jiashi Feng, Zheng Song, Tianzhu Zhang, Hanqing Lu, Changsheng Xu, and
  Shuicheng Yan.
\newblock Hi, magic closet, tell me what to wear!
\newblock In \emph{Proceedings of the 20th ACM international conference on
  Multimedia}, pages 619--628, 2012{\natexlab{a}}.

\bibitem[Liu et~al.(2012{\natexlab{b}})Liu, Song, Liu, Xu, Lu, and
  Yan]{liu2012street}
Si~Liu, Zheng Song, Guangcan Liu, Changsheng Xu, Hanqing Lu, and Shuicheng Yan.
\newblock Street-to-shop: Cross-scenario clothing retrieval via parts alignment
  and auxiliary set.
\newblock In \emph{2012 IEEE Conference on Computer Vision and Pattern
  Recognition}, pages 3330--3337. IEEE, 2012{\natexlab{b}}.

\bibitem[Liu et~al.(2013)Liu, Feng, Domokos, Xu, Huang, Hu, and
  Yan]{liu2013fashion}
Si~Liu, Jiashi Feng, Csaba Domokos, Hui Xu, Junshi Huang, Zhenzhen Hu, and
  Shuicheng Yan.
\newblock Fashion parsing with weak color-category labels.
\newblock \emph{IEEE Transactions on Multimedia}, 16\penalty0 (1):\penalty0
  253--265, 2013.

\bibitem[Liu et~al.(2014)Liu, Liu, and Yan]{liu2014fashion}
Si~Liu, Luoqi Liu, and Shuicheng Yan.
\newblock Fashion analysis: Current techniques and future directions.
\newblock \emph{IEEE MultiMedia}, 21\penalty0 (2):\penalty0 72--79, 2014.

\bibitem[Liu et~al.(2015)Liu, Liang, Liu, Lu, Lin, Cao, and
  Yan]{liu2015fashion}
Si~Liu, Xiaodan Liang, Luoqi Liu, Ke~Lu, Liang Lin, Xiaochun Cao, and Shuicheng
  Yan.
\newblock Fashion parsing with video context.
\newblock \emph{IEEE Transactions on Multimedia}, 17\penalty0 (8):\penalty0
  1347--1358, 2015.

\bibitem[Liu et~al.(2017)Liu, Wang, Qian, Yu, Bao, and
  Sun]{liu2017surveillance}
Si~Liu, Changhu Wang, Ruihe Qian, Han Yu, Renda Bao, and Yao Sun.
\newblock Surveillance video parsing with single frame supervision.
\newblock In \emph{Proceedings of the IEEE Conference on Computer Vision and
  Pattern Recognition}, pages 413--421, 2017.

\bibitem[Liu et~al.(2020)Liu, Sun, Liu, and Lin]{liu2020learning}
Xin Liu, Yongbin Sun, Ziwei Liu, and Dahua Lin.
\newblock Learning diverse fashion collocation by neural graph filtering.
\newblock \emph{arXiv preprint arXiv:2003.04888}, 2020.

\bibitem[Liu et~al.(2019{\natexlab{b}})Liu, Chen, Liu, and Lew]{liu2019swapgan}
Yu~Liu, Wei Chen, Li~Liu, and Michael~S Lew.
\newblock Swapgan: A multistage generative approach for person-to-person
  fashion style transfer.
\newblock \emph{IEEE Transactions on Multimedia}, 21\penalty0 (9):\penalty0
  2209--2222, 2019{\natexlab{b}}.

\bibitem[Liu et~al.(2016{\natexlab{b}})Liu, Luo, Qiu, Wang, and
  Tang]{liuLQWTcvpr16DeepFashion}
Ziwei Liu, Ping Luo, Shi Qiu, Xiaogang Wang, and Xiaoou Tang.
\newblock Deepfashion: Powering robust clothes recognition and retrieval with
  rich annotations.
\newblock In \emph{Proceedings of IEEE Conference on Computer Vision and
  Pattern Recognition (CVPR)}, June 2016{\natexlab{b}}.

\bibitem[Liu et~al.(2016{\natexlab{c}})Liu, Yan, Luo, Wang, and
  Tang]{liu2016fashion}
Ziwei Liu, Sijie Yan, Ping Luo, Xiaogang Wang, and Xiaoou Tang.
\newblock Fashion landmark detection in the wild.
\newblock In \emph{European Conference on Computer Vision}, pages 229--245.
  Springer, 2016{\natexlab{c}}.

\bibitem[Loni et~al.(2014)Loni, Cheung, Riegler, Bozzon, Gottlieb, and
  Larson]{loni2014fashion}
Babak Loni, Lei~Yen Cheung, Michael Riegler, Alessandro Bozzon, Luke Gottlieb,
  and Martha Larson.
\newblock Fashion 10000: an enriched social image dataset for fashion and
  clothing.
\newblock In \emph{Proceedings of the 5th ACM Multimedia Systems Conference},
  pages 41--46, 2014.

\bibitem[Lorbert et~al.(2017)Lorbert, Ben-Zvi, Ciptadi, Oks, and
  Tyagi]{lorbert2017toward}
Alexander Lorbert, Nir Ben-Zvi, Arridhana Ciptadi, Eduard Oks, and Ambrish
  Tyagi.
\newblock Toward better reconstruction of style images with gans.
\newblock \emph{KDDW on ML meets fashion}, 2017.

\bibitem[Ma et~al.(2018)Ma, Sun, Georgoulis, Van~Gool, Schiele, and
  Fritz]{ma2018disentangled}
Liqian Ma, Qianru Sun, Stamatios Georgoulis, Luc Van~Gool, Bernt Schiele, and
  Mario Fritz.
\newblock Disentangled person image generation.
\newblock In \emph{Proceedings of the IEEE Conference on Computer Vision and
  Pattern Recognition}, pages 99--108, 2018.

\bibitem[Ma et~al.(2019)Ma, Yang, Liao, Cao, and Chua]{ma2019and}
Yunshan Ma, Xun Yang, Lizi Liao, Yixin Cao, and Tat-Seng Chua.
\newblock Who, where, and what to wear? extracting fashion knowledge from
  social media.
\newblock In \emph{Proceedings of the 27th ACM International Conference on
  Multimedia}, pages 257--265, 2019.

\bibitem[Ma et~al.(2020)Ma, Ding, Yang, Liao, Wong, and Chua]{ma2020knowledge}
Yunshan Ma, Yujuan Ding, Xun Yang, Lizi Liao, Wai~Keung Wong, and Tat-Seng
  Chua.
\newblock Knowledge enhanced neural fashion trend forecasting.
\newblock In \emph{Proceedings of the 2020 International Conference on
  Multimedia Retrieval}, pages 82--90, 2020.

\bibitem[Mall et~al.(2020)Mall, Bala, Berg, and Grauman]{mall2020discovering}
Utkarsh Mall, Kavita Bala, Tamara Berg, and Kristen Grauman.
\newblock Discovering underground maps from fashion.
\newblock \emph{arXiv preprint arXiv:2012.02897}, 2020.

\bibitem[Manandhar et~al.(2018)Manandhar, Bastan, and Yap]{manandhar2018tiered}
Dipu Manandhar, Muhammet Bastan, and Kim-Hui Yap.
\newblock Tiered deep similarity search for fashion.
\newblock In \emph{Proceedings of the European Conference on Computer Vision
  (ECCV)}, pages 0--0, 2018.

\bibitem[MarketsandMarkets(2021)]{fashionreport}
MarketsandMarkets.
\newblock Ai in fashion market research report, 2021.
\newblock
  \url{https://www.reportlinker.com/p05953139/AI-in-Fashion-Market-Research-Report-by-Product-Type-by-Component-by-Deployment-by-Application-by-End-User-Global-Forecast-to-Cumulative-Impact-of-COVID-19.html?utm_source=GNW}.

\bibitem[Matzen et~al.(2017)Matzen, Bala, and Snavely]{StreetStyle2017}
Kevin Matzen, Kavita Bala, and Noah Snavely.
\newblock {StreetStyle}: {E}xploring world-wide clothing styles from millions
  of photos.
\newblock \emph{arXiv preprint arXiv:1706.01869}, 2017.

\bibitem[McAuley et~al.(2015)McAuley, Targett, Shi, and Van
  Den~Hengel]{mcauley2015image}
Julian McAuley, Christopher Targett, Qinfeng Shi, and Anton Van Den~Hengel.
\newblock Image-based recommendations on styles and substitutes.
\newblock In \emph{Proceedings of the 38th international ACM SIGIR conference
  on research and development in information retrieval}, pages 43--52, 2015.

\bibitem[Miura et~al.(2013)Miura, Yamasaki, and Aizawa]{miura2013snapper}
Shinya Miura, Toshihiko Yamasaki, and Kiyoharu Aizawa.
\newblock Snapper: fashion coordinate image retrieval system.
\newblock In \emph{2013 International Conference on Signal-Image Technology \&
  Internet-Based Systems}, pages 784--789. IEEE, 2013.

\bibitem[Natsume et~al.(2019)Natsume, Saito, Huang, Chen, Ma, Li, and
  Morishima]{natsume2019siclope}
Ryota Natsume, Shunsuke Saito, Zeng Huang, Weikai Chen, Chongyang Ma, Hao Li,
  and Shigeo Morishima.
\newblock Siclope: Silhouette-based clothed people.
\newblock In \emph{Proceedings of the IEEE/CVF Conference on Computer Vision
  and Pattern Recognition}, pages 4480--4490, 2019.

\bibitem[Pandey and Savakis(2020)]{pandey2020poly}
Nilesh Pandey and Andreas Savakis.
\newblock Poly-gan: Multi-conditioned gan for fashion synthesis.
\newblock \emph{Neurocomputing}, 414:\penalty0 356--364, 2020.

\bibitem[Patel et~al.(2020)Patel, Liao, and Pons-Moll]{patel2020tailornet}
Chaitanya Patel, Zhouyingcheng Liao, and Gerard Pons-Moll.
\newblock Tailornet: Predicting clothing in 3d as a function of human pose,
  shape and garment style.
\newblock In \emph{Proceedings of the IEEE/CVF Conference on Computer Vision
  and Pattern Recognition}, pages 7365--7375, 2020.

\bibitem[Polan{\'\i}a and Gupte(2019)]{polania2019learning}
Luisa~F Polan{\'\i}a and Satyajit Gupte.
\newblock Learning fashion compatibility across apparel categories for outfit
  recommendation.
\newblock In \emph{2019 IEEE International Conference on Image Processing
  (ICIP)}, pages 4489--4493. IEEE, 2019.

\bibitem[Reid(1850)]{reid1850essays}
Thomas Reid.
\newblock \emph{Essays on the intellectual powers of man}.
\newblock Number~66. J. Bartlett, 1850.

\bibitem[Sachdeva and Pandey(2020)]{sachdeva2020interactive}
Himani Sachdeva and Shreelekha Pandey.
\newblock Interactive systems for fashion clothing recommendation.
\newblock In \emph{Emerging Technology in Modelling and Graphics}, pages
  287--294. Springer, 2020.

\bibitem[Sagar et~al.(2020)Sagar, Garg, Kansal, Bhalla, Shah, and
  Yu]{sagar2020pai}
Dikshant Sagar, Jatin Garg, Prarthana Kansal, Sejal Bhalla, Rajiv~Ratn Shah,
  and Yi~Yu.
\newblock Pai-bpr: Personalized outfit recommendation scheme with
  attribute-wise interpretability.
\newblock In \emph{2020 IEEE Sixth International Conference on Multimedia Big
  Data (BigMM)}, pages 221--230. IEEE, 2020.

\bibitem[Saito et~al.(2019)Saito, Huang, Natsume, Morishima, Kanazawa, and
  Li]{saito2019pifu}
Shunsuke Saito, Zeng Huang, Ryota Natsume, Shigeo Morishima, Angjoo Kanazawa,
  and Hao Li.
\newblock Pifu: Pixel-aligned implicit function for high-resolution clothed
  human digitization.
\newblock In \emph{Proceedings of the IEEE/CVF International Conference on
  Computer Vision}, pages 2304--2314, 2019.

\bibitem[Sajja et~al.(2021)Sajja, Aggarwal, Mukherjee, Manglik, Dwivedi, and
  Raykar]{sajja2021explainable}
Shravan Sajja, Nupur Aggarwal, Sumanta Mukherjee, Kushagra Manglik, Satyam
  Dwivedi, and Vikas Raykar.
\newblock Explainable ai based interventions for pre-season decision making in
  fashion retail.
\newblock In \emph{8th ACM IKDD CODS and 26th COMAD}, pages 281--289. 2021.

\bibitem[Santesteban et~al.(2019)Santesteban, Otaduy, and
  Casas]{santesteban2019learning}
Igor Santesteban, Miguel~A Otaduy, and Dan Casas.
\newblock Learning-based animation of clothing for virtual try-on.
\newblock In \emph{Computer Graphics Forum}, volume~38, pages 355--366. Wiley
  Online Library, 2019.

\bibitem[Sbai et~al.(2018)Sbai, Elhoseiny, Bordes, LeCun, and
  Couprie]{sbai2018design}
Othman Sbai, Mohamed Elhoseiny, Antoine Bordes, Yann LeCun, and Camille
  Couprie.
\newblock Design: Design inspiration from generative networks.
\newblock In \emph{Proceedings of the European Conference on Computer Vision
  (ECCV) Workshops}, pages 0--0, 2018.

\bibitem[Shelley(2017)]{shelley2017concept}
James Shelley.
\newblock The concept of the aesthetic.
\newblock 2017.

\bibitem[Sherman et~al.(2019)Sherman, Shukla, Textor, Zhang, and
  Winecoff]{sherman2019assessing}
Jake Sherman, Chinmay Shukla, Rhonda Textor, Su~Zhang, and Amy~A Winecoff.
\newblock Assessing fashion recommendations: A multifaceted offline evaluation
  approach.
\newblock \emph{arXiv preprint arXiv:1909.04496}, 2019.

\bibitem[Shi et~al.(2019)Shi, Hui, Liu, Lin, and Loy]{shi2019learning}
Wu~Shi, Tak-Wai Hui, Ziwei Liu, Dahua Lin, and Chen~Change Loy.
\newblock Learning to synthesize fashion textures.
\newblock \emph{arXiv preprint arXiv:1911.07472}, 2019.

\bibitem[Siarohin et~al.(2018)Siarohin, Sangineto, Lathuiliere, and
  Sebe]{siarohin2018deformable}
Aliaksandr Siarohin, Enver Sangineto, St{\'e}phane Lathuiliere, and Nicu Sebe.
\newblock Deformable gans for pose-based human image generation.
\newblock In \emph{Proceedings of the IEEE Conference on Computer Vision and
  Pattern Recognition}, pages 3408--3416, 2018.

\bibitem[Simo-Serra et~al.(2015)Simo-Serra, Fidler, Moreno-Noguer, and
  Urtasun]{SimoSerraCVPR2015}
Edgar Simo-Serra, Sanja Fidler, Francesc Moreno-Noguer, and Raquel Urtasun.
\newblock {Neuroaesthetics in Fashion: Modeling the Perception of
  Fashionability}.
\newblock In \emph{Proceedings of the Conference on Computer Vision and Pattern
  Recognition (CVPR)}, 2015.

\bibitem[Singh et~al.(2019)Singh, Gupta, Jha, and Rajan]{singh2019fashion}
Pawan~Kumar Singh, Yadunath Gupta, Nilpa Jha, and Aruna Rajan.
\newblock Fashion retail: Forecasting demand for new items.
\newblock \emph{arXiv preprint arXiv:1907.01960}, 2019.

\bibitem[Singhal et~al.(2020)Singhal, Chopra, Ayush, Govind, and
  Krishnamurthy]{singhal2020towards}
Anirudh Singhal, Ayush Chopra, Kumar Ayush, Utkarsh~Patel Govind, and Balaji
  Krishnamurthy.
\newblock Towards a unified framework for visual compatibility prediction.
\newblock In \emph{Proceedings of the IEEE/CVF Winter Conference on
  Applications of Computer Vision}, pages 3607--3616, 2020.

\bibitem[Song and Mei(2018)]{song2018multimedia}
Sijie Song and Tao Mei.
\newblock When multimedia meets fashion.
\newblock \emph{IEEE MultiMedia}, 25\penalty0 (3):\penalty0 102--108, 2018.

\bibitem[Song et~al.(2017)Song, Feng, Liu, Li, Nie, and
  Ma]{song2017neurostylist}
Xuemeng Song, Fuli Feng, Jinhuan Liu, Zekun Li, Liqiang Nie, and Jun Ma.
\newblock Neurostylist: Neural compatibility modeling for clothing matching.
\newblock In \emph{Proceedings of the 25th ACM international conference on
  Multimedia}, pages 753--761, 2017.

\bibitem[Song et~al.(2018)Song, Feng, Han, Yang, Liu, and Nie]{song2018neural}
Xuemeng Song, Fuli Feng, Xianjing Han, Xin Yang, Wei Liu, and Liqiang Nie.
\newblock Neural compatibility modeling with attentive knowledge distillation.
\newblock In \emph{The 41st International ACM SIGIR Conference on Research \&
  Development in Information Retrieval}, pages 5--14, 2018.

\bibitem[Song et~al.(2019)Song, Han, Li, Chen, Xu, and Nie]{song2019gp}
Xuemeng Song, Xianjing Han, Yunkai Li, Jingyuan Chen, Xin-Shun Xu, and Liqiang
  Nie.
\newblock Gp-bpr: Personalized compatibility modeling for clothing matching.
\newblock In \emph{Proceedings of the 27th ACM International Conference on
  Multimedia}, pages 320--328, 2019.

\bibitem[Sonie et~al.(2019)Sonie, Chelliah, and Sural]{sonie2019personalised}
Omprakash Sonie, Muthusamy Chelliah, and Shamik Sural.
\newblock Personalised fashion recommendation using deep learning.
\newblock In \emph{Proceedings of the ACM India Joint International Conference
  on Data Science and Management of Data}, pages 368--368, 2019.

\bibitem[Stefani et~al.(2019)Stefani, Stefanis, and
  Garofalakis]{stefani2019cfrs}
Maria~Anastassia Stefani, Vassilios Stefanis, and John Garofalakis.
\newblock Cfrs: A trends-driven collaborative fashion recommendation system.
\newblock In \emph{2019 10th International Conference on Information,
  Intelligence, Systems and Applications (IISA)}, pages 1--4. IEEE, 2019.

\bibitem[Su et~al.(2020)Su, Yu, Wang, Li, and Liu]{su2020deepcloth}
Zhaoqi Su, Tao Yu, Yangang Wang, Yipeng Li, and Yebin Liu.
\newblock Deepcloth: Neural garment representation for shape and style editing.
\newblock \emph{arXiv preprint arXiv:2011.14619}, 2020.

\bibitem[Sun et~al.(2020)Sun, He, Wu, Zhao, and Peng]{sun2020learning}
Guang-Lu Sun, Jun-Yan He, Xiao Wu, Bo~Zhao, and Qiang Peng.
\newblock Learning fashion compatibility across categories with deep multimodal
  neural networks.
\newblock \emph{Neurocomputing}, 395:\penalty0 237--246, 2020.

\bibitem[Sung et~al.(2020)Sung, Baek, Sim, Kim, Hwangbo, and
  Jang]{sung2020breaking}
Shin~Woong Sung, Hyunsuk Baek, Hyeonjun Sim, Eun~Hie Kim, Hyunwoo Hwangbo, and
  Young~Jae Jang.
\newblock Breaking moravec's paradox: Visual-based distribution in smart
  fashion retail.
\newblock \emph{arXiv preprint arXiv:2007.09102}, 2020.

\bibitem[Takagi et~al.(2017)Takagi, Simo-Serra, Iizuka, and
  Ishikawa]{takagi2017makes}
Moeko Takagi, Edgar Simo-Serra, Satoshi Iizuka, and Hiroshi Ishikawa.
\newblock What makes a style: Experimental analysis of fashion prediction.
\newblock In \emph{Proceedings of the IEEE International Conference on Computer
  Vision Workshops}, pages 2247--2253, 2017.

\bibitem[Takahashi et~al.(2020)Takahashi, Yamaguchi, and
  Watanabe]{takahashi2020cat}
Satoshi Takahashi, Keiko Yamaguchi, and Asuka Watanabe.
\newblock Cat street: Chronicle archive of tokyo street-fashion.
\newblock \emph{arXiv preprint arXiv:2009.13395}, 2020.

\bibitem[Tango et~al.(2020)Tango, Katsurai, Maki, and Goto]{tango2020anime}
Koya Tango, Marie Katsurai, Hayato Maki, and Ryosuke Goto.
\newblock Anime-to-real clothing: Cosplay costume generation via image-to-image
  translation.
\newblock \emph{arXiv preprint arXiv:2008.11479}, 2020.

\bibitem[Tangseng and Okatani(2020)]{tangseng2020toward}
Pongsate Tangseng and Takayuki Okatani.
\newblock Toward explainable fashion recommendation.
\newblock In \emph{Proceedings of the IEEE/CVF Winter Conference on
  Applications of Computer Vision}, pages 2153--2162, 2020.

\bibitem[Tangseng et~al.(2017{\natexlab{a}})Tangseng, Wu, and
  Yamaguchi]{tangseng2017looking}
Pongsate Tangseng, Zhipeng Wu, and Kota Yamaguchi.
\newblock Looking at outfit to parse clothing.
\newblock \emph{arXiv preprint arXiv:1703.01386}, 2017{\natexlab{a}}.

\bibitem[Tangseng et~al.(2017{\natexlab{b}})Tangseng, Yamaguchi, and
  Okatani]{tangseng2017recommending}
Pongsate Tangseng, Kota Yamaguchi, and Takayuki Okatani.
\newblock Recommending outfits from personal closet.
\newblock In \emph{Proceedings of the IEEE International Conference on Computer
  Vision Workshops}, pages 2275--2279, 2017{\natexlab{b}}.

\bibitem[Tian et~al.(2021)Tian, Chanda, Kumar, and Gray]{tian2021improving}
Qing Tian, Sampath Chanda, KC~Kumar, and Douglas Gray.
\newblock Improving apparel detection with category grouping and multi-grained
  branches.
\newblock \emph{arXiv preprint arXiv:2101.06770}, 2021.

\bibitem[Varol et~al.(2018)Varol, Ceylan, Russell, Yang, Yumer, Laptev, and
  Schmid]{varol2018bodynet}
Gul Varol, Duygu Ceylan, Bryan Russell, Jimei Yang, Ersin Yumer, Ivan Laptev,
  and Cordelia Schmid.
\newblock Bodynet: Volumetric inference of 3d human body shapes.
\newblock In \emph{Proceedings of the European Conference on Computer Vision
  (ECCV)}, pages 20--36, 2018.

\bibitem[Vashishtha et~al.(2020)Vashishtha, Burman, Kumar, Sethuraman, Sekar,
  and Ramanan]{vashishtha2020product}
Rajesh~Kumar Vashishtha, Vibhati Burman, Rajan Kumar, Srividhya Sethuraman,
  Abhinaya~R Sekar, and Sharadha Ramanan.
\newblock Product age based demand forecast model for fashion retail.
\newblock \emph{arXiv preprint arXiv:2007.05278}, 2020.

\bibitem[Vasileva et~al.(2018)Vasileva, Plummer, Dusad, Rajpal, Kumar, and
  Forsyth]{VasilevaECCV18FasionCompatibility}
Mariya~I. Vasileva, Bryan~A. Plummer, Krishna Dusad, Shreya Rajpal, Ranjitha
  Kumar, and David Forsyth.
\newblock Learning type-aware embeddings for fashion compatibility.
\newblock In \emph{ECCV}, 2018.

\bibitem[Verma et~al.(2020)Verma, Gulati, and Shah]{verma2020addressing}
Dhruv Verma, Kshitij Gulati, and Rajiv~Ratn Shah.
\newblock Addressing the cold-start problem in outfit recommendation using
  visual preference modelling.
\newblock In \emph{2020 IEEE Sixth International Conference on Multimedia Big
  Data (BigMM)}, pages 251--256. IEEE, 2020.

\bibitem[Vittayakorn et~al.(2016)Vittayakorn, Umeda, Murasaki, Sudo, Okatani,
  and Yamaguchi]{vittayakorn2016automatic}
Sirion Vittayakorn, Takayuki Umeda, Kazuhiko Murasaki, Kyoko Sudo, Takayuki
  Okatani, and Kota Yamaguchi.
\newblock Automatic attribute discovery with neural activations.
\newblock In \emph{European Conference on Computer Vision}, pages 252--268.
  Springer, 2016.

\bibitem[Wang and Ai(2011)]{wang2011blocks}
Nan Wang and Haizhou Ai.
\newblock Who blocks who: Simultaneous clothing segmentation for grouping
  images.
\newblock In \emph{2011 International Conference on Computer Vision}, pages
  1535--1542. IEEE, 2011.

\bibitem[Wang et~al.(2019)Wang, Wu, and Zhong]{wang2019outfit}
Xin Wang, Bo~Wu, and Yueqi Zhong.
\newblock Outfit compatibility prediction and diagnosis with multi-layered
  comparison network.
\newblock In \emph{Proceedings of the 27th ACM International Conference on
  Multimedia}, pages 329--337, 2019.

\bibitem[Wu et~al.(2021)Wu, Chao, Chen, Xu, Liu, Manocha, Sun, Han, Yao, and
  Jin]{wu2021example}
Nannan Wu, Qianwen Chao, Yanzhen Chen, Weiwei Xu, Chen Liu, Dinesh Manocha,
  Wenxin Sun, Yi~Han, Xinran Yao, and Xiaogang Jin.
\newblock Example-based real-time clothing synthesis for virtual agents.
\newblock \emph{arXiv preprint arXiv:2101.03088}, 2021.

\bibitem[Wu and Boulanger(2016)]{wu2016enhanced}
Qiong Wu and Pierre Boulanger.
\newblock Enhanced reweighted mrfs for efficient fashion image parsing.
\newblock \emph{ACM Transactions on Multimedia Computing, Communications, and
  Applications (TOMM)}, 12\penalty0 (3):\penalty0 1--16, 2016.

\bibitem[Xia et~al.(2017)Xia, Wang, Chen, and Yuille]{xia2017joint}
Fangting Xia, Peng Wang, Xianjie Chen, and Alan~L Yuille.
\newblock Joint multi-person pose estimation and semantic part segmentation.
\newblock In \emph{Proceedings of the IEEE conference on computer vision and
  pattern recognition}, pages 6769--6778, 2017.

\bibitem[Xian et~al.(2018)Xian, Sangkloy, Agrawal, Raj, Lu, Fang, Yu, and
  Hays]{xian2018texturegan}
Wenqi Xian, Patsorn Sangkloy, Varun Agrawal, Amit Raj, Jingwan Lu, Chen Fang,
  Fisher Yu, and James Hays.
\newblock Texturegan: Controlling deep image synthesis with texture patches.
\newblock In \emph{Proceedings of the IEEE Conference on Computer Vision and
  Pattern Recognition}, pages 8456--8465, 2018.

\bibitem[Xiao et~al.(2015)Xiao, Xia, Yang, Huang, and Wang]{xiao2015learning}
Tong Xiao, Tian Xia, Yi~Yang, Chang Huang, and Xiaogang Wang.
\newblock Learning from massive noisy labeled data for image classification.
\newblock In \emph{CVPR}, 2015.

\bibitem[Yamaguchi et~al.(2012)Yamaguchi, Kiapour, Ortiz, and
  Berg]{yamaguchi2012parsing}
Kota Yamaguchi, M~Hadi Kiapour, Luis~E Ortiz, and Tamara~L Berg.
\newblock Parsing clothing in fashion photographs.
\newblock In \emph{2012 IEEE Conference on Computer vision and pattern
  recognition}, pages 3570--3577. IEEE, 2012.

\bibitem[Yamaguchi et~al.(2013)Yamaguchi, Hadi~Kiapour, and
  Berg]{yamaguchi2013paper}
Kota Yamaguchi, M~Hadi~Kiapour, and Tamara~L Berg.
\newblock Paper doll parsing: Retrieving similar styles to parse clothing
  items.
\newblock In \emph{Proceedings of the IEEE international conference on computer
  vision}, pages 3519--3526, 2013.

\bibitem[Yamaguchi et~al.(2014)Yamaguchi, Kiapour, Ortiz, and
  Berg]{yamaguchi2014retrieving}
Kota Yamaguchi, M~Hadi Kiapour, Luis~E Ortiz, and Tamara~L Berg.
\newblock Retrieving similar styles to parse clothing.
\newblock \emph{IEEE transactions on pattern analysis and machine
  intelligence}, 37\penalty0 (5):\penalty0 1028--1040, 2014.

\bibitem[Yamaguchi et~al.(2015)Yamaguchi, Okatani, Sudo, Murasaki, and
  Taniguchi]{yamaguchi2015mix}
Kota Yamaguchi, Takayuki Okatani, Kyoko Sudo, Kazuhiko Murasaki, and Yukinobu
  Taniguchi.
\newblock Mix and match: Joint model for clothing and attribute recognition.
\newblock In \emph{BMVC}, volume~1, page~4, 2015.

\bibitem[Yan et~al.(2017)Yan, Liu, Luo, Qiu, Wang, and
  Tang]{yan2017unconstrained}
Sijie Yan, Ziwei Liu, Ping Luo, Shi Qiu, Xiaogang Wang, and Xiaoou Tang.
\newblock Unconstrained fashion landmark detection via hierarchical recurrent
  transformer networks.
\newblock In \emph{Proceedings of the 25th ACM international conference on
  Multimedia}, pages 172--180, 2017.

\bibitem[Yang et~al.(2019)Yang, He, Wang, Ma, Feng, Wang, and
  Chua]{yang2019interpretable}
Xun Yang, Xiangnan He, Xiang Wang, Yunshan Ma, Fuli Feng, Meng Wang, and
  Tat-Seng Chua.
\newblock Interpretable fashion matching with rich attributes.
\newblock In \emph{Proceedings of the 42nd International ACM SIGIR Conference
  on Research and Development in Information Retrieval}, pages 775--784, 2019.

\bibitem[Yang et~al.(2020)Yang, Du, and Wang]{yang2020learning}
Xun Yang, Xiaoyu Du, and Meng Wang.
\newblock Learning to match on graph for fashion compatibility modeling.
\newblock In \emph{Proceedings of the AAAI Conference on Artificial
  Intelligence}, volume~34, pages 287--294, 2020.

\bibitem[Yoon et~al.(2021)Yoon, Kim, Kautz, and Park]{yoon2021neural}
Jae~Shin Yoon, Kihwan Kim, Jan Kautz, and Hyun~Soo Park.
\newblock Neural 3d clothes retargeting from a single image.
\newblock \emph{arXiv preprint arXiv:2102.00062}, 2021.

\bibitem[Yu and Grauman(2014{\natexlab{a}})]{finegrained}
A.~Yu and K.~Grauman.
\newblock Fine-grained visual comparisons with local learning.
\newblock In \emph{Computer Vision and Pattern Recognition (CVPR)}, Jun
  2014{\natexlab{a}}.

\bibitem[Yu and Grauman(2014{\natexlab{b}})]{yu2014fine}
Aron Yu and Kristen Grauman.
\newblock Fine-grained visual comparisons with local learning.
\newblock In \emph{Proceedings of the IEEE Conference on Computer Vision and
  Pattern Recognition}, pages 192--199, 2014{\natexlab{b}}.

\bibitem[Yu et~al.(2019{\natexlab{a}})Yu, Hu, Chen, and
  Zeng]{yu2019personalized}
Cong Yu, Yang Hu, Yan Chen, and Bing Zeng.
\newblock Personalized fashion design.
\newblock In \emph{Proceedings of the IEEE/CVF International Conference on
  Computer Vision}, pages 9046--9055, 2019{\natexlab{a}}.

\bibitem[Yu et~al.(2019{\natexlab{b}})Yu, Liang, Gong, Jiang, Xiao, and
  Lin]{yu2019layout}
Weijiang Yu, Xiaodan Liang, Ke~Gong, Chenhan Jiang, Nong Xiao, and Liang Lin.
\newblock Layout-graph reasoning for fashion landmark detection.
\newblock In \emph{Proceedings of the IEEE/CVF Conference on Computer Vision
  and Pattern Recognition}, pages 2937--2945, 2019{\natexlab{b}}.

\bibitem[Yuan and Moghaddam(2020)]{yuan2020garment}
Chenxi Yuan and Mohsen Moghaddam.
\newblock Garment design with generative adversarial networks.
\newblock \emph{arXiv preprint arXiv:2007.10947}, 2020.

\bibitem[Zhan et~al.(2017{\natexlab{a}})Zhan, Shi, and Kot]{zhan2017cross}
Huijing Zhan, Boxin Shi, and Alex~C Kot.
\newblock Cross-domain shoe retrieval with a semantic hierarchy of attribute
  classification network.
\newblock \emph{IEEE Transactions on Image Processing}, 26\penalty0
  (12):\penalty0 5867--5881, 2017{\natexlab{a}}.

\bibitem[Zhan et~al.(2017{\natexlab{b}})Zhan, Shi, and Kot]{zhan2017street}
Huijing Zhan, Boxin Shi, and Alex~C Kot.
\newblock Street-to-shop shoe retrieval with multi-scale viewpoint invariant
  triplet network.
\newblock In \emph{2017 IEEE International Conference on Image Processing
  (ICIP)}, pages 1102--1106. IEEE, 2017{\natexlab{b}}.

\bibitem[Zhan et~al.(2019)Zhan, Shi, Duan, and Kot]{zhan2019deepshoe}
Huijing Zhan, Boxin Shi, Ling-Yu Duan, and Alex~C Kot.
\newblock Deepshoe: An improved multi-task view-invariant cnn for
  street-to-shop shoe retrieval.
\newblock \emph{Computer Vision And Image Understanding}, 180:\penalty0 23--33,
  2019.

\bibitem[Zhan et~al.(2020)Zhan, Yi, Shi, Lin, Duan, and Kot]{zhan2020pose}
Huijing Zhan, Chenyu Yi, Boxin Shi, Jie Lin, Ling-Yu Duan, and Alex~C Kot.
\newblock Pose-normalized and appearance-preserved street-to-shop clothing
  image generation and feature learning.
\newblock \emph{IEEE Transactions on Multimedia}, 23:\penalty0 133--144, 2020.

\bibitem[Zhang et~al.(2018)Zhang, Huang, Liu, and Xu]{zhang2018clothes}
Haijun Zhang, Wang Huang, Linlin Liu, and Xiaofei Xu.
\newblock Clothes collocation recommendations by compatibility learning.
\newblock In \emph{2018 IEEE International Conference on Web Services (ICWS)},
  pages 179--186. IEEE, 2018.

\bibitem[Zhang et~al.(2020{\natexlab{a}})Zhang, Yang, Tan, Wu, Wang, and
  Kuo]{zhang2020learning}
Heming Zhang, Xuewen Yang, Jianchao Tan, Chi-Hao Wu, Jue Wang, and C-C~Jay Kuo.
\newblock Learning color compatibility in fashion outfits.
\newblock \emph{arXiv preprint arXiv:2007.02388}, 2020{\natexlab{a}}.

\bibitem[Zhang et~al.(2020{\natexlab{b}})Zhang, Wang, Ceylan, and
  Mitra]{zhang2020deep}
Meng Zhang, Tuanfeng Wang, Duygu Ceylan, and Niloy~J Mitra.
\newblock Deep detail enhancement for any garment.
\newblock \emph{arXiv e-prints}, pages arXiv--2008, 2020{\natexlab{b}}.

\bibitem[Zhao and Ma(2018)]{zhao2018compensation}
Zhenjie Zhao and Xiaojuan Ma.
\newblock A compensation method of two-stage image generation for human-ai
  collaborated in-situ fashion design in augmented reality environment.
\newblock In \emph{2018 IEEE International Conference on Artificial
  Intelligence and Virtual Reality (AIVR)}, pages 76--83. IEEE, 2018.

\bibitem[Zheng et~al.(2020)Zheng, Wu, Park, Zhu, and
  Luo]{zheng2020personalized}
Haitian Zheng, Kefei Wu, Jong-Hwi Park, Wei Zhu, and Jiebo Luo.
\newblock Personalized fashion recommendation from personal social media data:
  An item-to-set metric learning approach.
\newblock \emph{arXiv preprint arXiv:2005.12439}, 2020.

\bibitem[Zhou et~al.(2019)Zhou, Huang, Li, Li, Li, and Zhang]{zhou2019text}
Xingran Zhou, Siyu Huang, Bin Li, Yingming Li, Jiachen Li, and Zhongfei Zhang.
\newblock Text guided person image synthesis.
\newblock In \emph{Proceedings of the IEEE/CVF Conference on Computer Vision
  and Pattern Recognition}, pages 3663--3672, 2019.

\bibitem[Zhu et~al.(2017)Zhu, Urtasun, Fidler, Lin, and
  Change~Loy]{zhu2017your}
Shizhan Zhu, Raquel Urtasun, Sanja Fidler, Dahua Lin, and Chen Change~Loy.
\newblock Be your own prada: Fashion synthesis with structural coherence.
\newblock In \emph{Proceedings of the IEEE international conference on computer
  vision}, pages 1680--1688, 2017.

\bibitem[Ziegler et~al.(2020)Ziegler, Butepage, Welle, Varava, Novkovic, and
  Kragic]{ziegler2020fashion}
Thomas Ziegler, Judith Butepage, Michael~C Welle, Anastasiia Varava, Tonci
  Novkovic, and Danica Kragic.
\newblock Fashion landmark detection and category classification for robotics.
\newblock In \emph{2020 IEEE International Conference on Autonomous Robot
  Systems and Competitions (ICARSC)}, pages 81--88. IEEE, 2020.

\bibitem[Zou et~al.(2018)Zou, Wong, and Mo]{zou2018fashion}
Xingxing Zou, Wai~Keung Wong, and Dongmei Mo.
\newblock Fashion meets ai technology.
\newblock In \emph{International Conference on Artificial Intelligence on
  Textile and Apparel}, pages 255--267. Springer, 2018.

\bibitem[Zou et~al.(2019)Zou, Kong, Wong, Wang, Liu, and Cao]{zou2019fashionai}
Xingxing Zou, Xiangheng Kong, Waikeung Wong, Congde Wang, Yuguang Liu, and Yang
  Cao.
\newblock Fashionai: A hierarchical dataset for fashion understanding.
\newblock In \emph{Proceedings of the IEEE Conference on Computer Vision and
  Pattern Recognition Workshops}, pages 0--0, 2019.

\bibitem[Zou et~al.(2020)Zou, Li, Bai, Lin, and Wong]{zou2020regularizing}
Xingxing Zou, Zhizhong Li, Ke~Bai, Dahua Lin, and Waikeung Wong.
\newblock Regularizing reasons for outfit evaluation with gradient penalty.
\newblock \emph{arXiv preprint arXiv:2002.00460}, 2020.

\end{thebibliography}
\end{document}